\documentstyle[11pt,aaspp4]{article}

\newcommand{\msun}{M_{\odot}}

\newcommand{\zsun}{Z_{\odot}}
\newcommand{\kms}{\, {\rm km\, s}^{-1}}
\newcommand{\cm}{\, {\rm cm}}
\newcommand{\erg}{\, {\rm erg}}
\newcommand{\kpc}{\, {\rm Kpc}}
\newcommand{\mpc}{\, {\rm Mpc}}

\newcommand{\lya}{Ly$\alpha$ }
\newcommand{\etal}{et al.\ }

\newcommand{\rhoc}{{\rho_{\rm c}}}
\newcommand{\nhi}{N_{\rm HI}}
\newcommand{\eq}{equation }

\begin{document}

\title{Galaxy Formation and the Kinematics of Damped \lya Systems}

\author{Patrick McDonald and Jordi Miralda-Escud\'e$^{1}$}
\affil{ Univ. of Pennsylvania, Dept. of Physics and Astronomy}
\authoremail{ pmcdonal@student.physics.upenn.edu, 
              jordi@llull.physics.upenn.edu}
\authoraddr{David Rittenhouse Laboratory, 209 S. 33rd St., 
            Philadelphia, PA 19104}            

\affil{$^{1}$ Alfred P. Sloan Fellow}

\begin{abstract}

A model of damped \lya systems is presented based on randomly moving clouds in
spherical halos. We use the Press-Schechter model for the abundance of halos,
and assume that each halo has a similar population of clouds, with total mass
and spatial distribution constrained to fit observations of the column density
distribution. We show that the kinematics of the multiple absorbing components
revealed in absorption profiles of the low-ionization lines, presented by
Prochaska \& Wolfe, are consistent with our spherical halo model. 

The presence of multiple absorbing components with a large covering factor,
combined with the small impact parameters of the systems predicted in our
analytical model and in numerical simulations, implies a high rate of energy
dissipation in cloud collisions. We calculate the rate of energy dissipation in
our model, and show that it is far greater than the rate at which energy can be
supplied by gravitational mergers of halos. This poses a problem for the model
of merging protogalactic clumps of Haehnelt et al., based on numerical
simulations. We also present new constraints on the amplitude of the power
spectrum in hierarchical theories required to account for the observed
velocity dispersion in the absorbers. We find that the linearly extrapolated
rms fluctuation on spheres of radius $HR = 100 \kms$ at $z=4$ must be 
greater than 0.75. 
Although this limit is obtained only for our specific model of the
absorbing components, it should not be highly model-dependent because
the velocity dispersion of the absorbers is essentially determined by the
velocity dispersion of the halos where the gas is moving.

\end{abstract}


\section{Introduction}

  The highest column density \lya absorption systems observed in 
quasar
spectra offer a powerful technique to investigate the structure and
evolution of high density clouds of neutral hydrogen that must have
been present at the sites where galaxies formed. Observations show
that $\sim 20\%$ of the lines of sight to high-redshift quasars
($z\sim 3$) contain an absorption system with column density
$\nhi > 2\times 10^{20}\cm ^{-2}$ (the generally adopted column
density threshold for classifying an absorber as a ``damped''
system), with the number per unit redshift increasing slowly with
$z$, and probably decreasing at $z > 3$ for high column densities
(Wolfe \etal 1993; Storrie-Lombardi \etal 1996). These systems
contain most of the neutral hydrogen in the universe, although most
of the ionized hydrogen is probably in lower density structures.
The total neutral hydrogen contained in damped \lya systems at high
redshift is similar to the mass presently contained in stars, and
accounts for $\sim 10\%$ of the baryonic density predicted from
primordial nucleosynthesis (Storrie-Lombardi \etal 1996; 
Lanzetta \etal 1991).  The abundance of heavy elements
is only $Z\sim 10^{-2} \zsun$ at $z\sim 3$, and increases at lower
redshifts to a mean of $Z\sim 0.1 \zsun$ at $z=1$, although with a
large scatter (Lu \etal 1996; Pettini \etal 1994). 
The observational evidence
on the physical size of the absorbing systems is still scant; a few
observations of associated $21 \cm$ absorption or \lya emission
indicate overall sizes of the systems of $\sim 10 h^{-1} \kpc$ or
larger, but with significant clumpiness on smaller scales
(Briggs \etal 1989; Wolfe \etal 1992; Djorgovski \etal 1996;
Djorgovski 1997;
but see also Moller \& Warren 1998).

  The generic framework whereby these observations have been
interpreted in recent years is that of the models of structure formation
by gravitational collapse of primordial fluctuations. Although a theory
of the origin of these fluctuations that can predict the power spectrum
does not yet exist (the various cosmological models normally used
are based on the assumption that a hypothetical process, such as the
quantum fluctuations of a field driving inflation, gives rise to
perfectly scale-invariant and Gaussian fluctuations), a large body of
observational evidence has accumulated pointing to a hierarchical
theory where dark matter halos collapse from primordial fluctuations,
starting on small scales at high redshifts and merging on larger scales
later. This hierarchical scenario is a generic consequence of the
presence of any type of dark matter than can collapse on small scales
(i.e., ``cold'' dark matter), with a power spectrum of density
fluctuations that does not
abruptly decrease at some special ``smoothing'' scale.
The evidence for models of this type includes the observed hierarchical
nature of galaxy clustering and the galaxy peculiar velocity field
(e.g., Strauss \& Willick 1995), the presence of substructure in galaxy
clusters indicating that clusters formed recently from mergers of
smaller units (e.g., Bird 1994), the \lya forest at high redshift
showing the structures on small scales that were collapsing in the past
(Rauch \etal 1997 and references therein), and the CMB fluctuations
(White, Scott, \& Silk 1994).

  The formation of galaxies in a hierarchical picture occurs in dark
matter halos where the cooling time of the gas is sufficiently short.
The gas can then dissipate its energy and concentrate in the centers
of halos (White \& Rees 1978). In general, halos where galaxies can
form have velocity dispersions approximately in the range $30 - 300
\kms$. In smaller halos, the temperature of the photoionized gas is
sufficiently high to prevent collapse and efficient cooling (e.g.,
Thoul \& Weinberg 1996; Navarro \& Steinmetz 1997).
Larger halos have a cooling time that is too long at the time when 
they collapse, so they contain clusters of preexisting galaxies
(e.g., Ostriker \& Rees 1977). 

  Analytic models and cosmological simulations have shown that
the observed amount of gas in damped \lya systems can originate
from the cooled gas in halos of galactic scale. 
The fraction of baryons that collapse in halos and can efficiently
cool can account for the observed mass in damped \lya systems in most
models (Mo \& Miralda-Escud\'e 1994, Kauffmann \& Charlot 1994,
Ma \& Bertschinger 1994, Katz \etal 1996, Gardner \etal 1997a,b),
as long as a large fraction
of the collapsed gas in these halos can remain in atomic form
(requiring the gas to be dense enough to be shielded from the
ionizing background, but not turning rapidly to molecular gas or stars).
The low metallicities
of damped \lya systems at $z\sim 3$ (Lu \etal 1996)
also suggest that most of the gas that dissipates
in halos at high redshift does not quickly turn to stars,
but remains atomic for a long time. The gas may form
rotating disks, as in spiral galaxies, or may also be distributed more
irregularly in the halo as a consequence of frequent merging
(see Haehnelt, Steinmetz, \& Rauch 1998).

  The fundamental physical process that governs the formation of
galaxies is the dissipation of gas in dark matter halos. The absorption
systems offer us an excellent way to observe the gas that is undergoing
this process. Several key questions connecting the observations and the
theory emerge here:
 What is the column density of the absorption
systems containing the gas in the process of dissipating most of the
energy that needs to be lost to form galaxies? 
Are the damped \lya systems the site of most of the dissipation,
or are they stable rotating disks where gas
accumulates after having already dissipated its energy as a lower column
density system?
Can we measure the rate of dissipation from
observations? Theoretical models  
of the damped absorbers have so far focused on the total observed amount
of neutral hydrogen and its distribution in column density as a tool
to constrain cosmological models of structure formation; but can we also
test whether the neutral hydrogen is located in halos with the predicted
distribution of velocity dispersions? 

  The answer to these questions requires kinematic studies of the
damped \lya systems. Recently, Prochaska \& Wolfe (1997, 
hereafter PW1) presented
an analysis of high-resolution observations of 17 metal line 
absorption systems associated with damped \lya systems. 
Prochaska \& Wolfe (1998, hereafter PW2) expanded the sample to a 
total of 31 systems.
The velocity structure of the
gas can be inferred from the unsaturated absorption profiles of
low-ionization species, which can be reasonably assumed to have a
constant fractional abundance throughout the self-shielded region of
the absorption system. In this paper we shall construct a simple model
of the multiple absorbers in the damped \lya systems, based on a
spherical distribution of halo clouds,
to interpret these observations and to relate them to the questions
we have asked above. Our model is presented in \S 2.
In \S 3 we use the observations of the kinematics of damped \lya
systems to estimate the rate at which energy should be dissipated;
we also present an analytic model to approximately compute the amount
of energy that can be gained by the absorbing gas owing to the collapse
and mergers of galaxies in dark matter halos. The comparison of our
model predictions to observations is done in \S 4.

\section{Model}

Damped \lya systems are often interpreted as large rotating disks
of gas that are the progenitors of disk galaxies (e.g., Wolfe 1995).
Although it is indeed expected that rotating disks will form in halos
of galactic scales (following rapid dissipation with conservation of
angular momentum; e.g., Fall \& Efstathiou 1980), obviously the gas 
must
first move through the halo in order to form disks. If the dissipation
process takes a substantial time, much of the observed gas in damped
systems at high redshift could still be undergoing
dissipation at a large radius in halos,
where rotational support is not important. In fact, if we assume that
most of the baryons in a spherical halo with circular velocity $V_c$ are
in the form of neutral hydrogen (self-shielded from the external
ionizing background), accounting for a fraction $f_b$ of the total
halo mass, with a density profile $\rho_{HI} \propto r^{-2}$,
the column density at impact parameter $b$ is
$N_{HI} = 5.5\times 10^{21}\, f_b\, (V_c/100 \kms)\, h\,
[(1+z)/4]^{3/2}\, (r_{vir}/b)\, \cm^{-2}$,
where $r_{vir} = V_c\, t/(2\pi)$ is the virial radius, defined as the
radius containing a mean density $178 \,\rho_{crit}$ 
(we use $h=H_0/(100\kms\mpc^{-1})$, and
$t=2/(3H_0) (1+z)^{-3/2}$, for the case $\Omega=1$).
Clearly, for a baryon fraction $f_b \sim 0.1$,
the column densities of damped systems ($N_{HI} > 2\times 10^{20}
\cm^{-2}$) can be reached even at the outermost regions of virialized
halos if the baryons are not very centrally concentrated relative to the
dark matter.

  The possibility that the damped \lya systems are largely associated
with halo gas clouds has been proposed by Haehnelt \etal (1998),
based on hydrodynamic cosmological simulations where it
is indeed found that much of the gas in halos has not had
time to settle to a rotating configuration in equilibrium, owing to the
high frequency of mergers. Here, we shall study a simple
analytic model to understand the origin of this result in terms of
the rate at which energy can be dissipated, and to relate the observed
kinematics to predictions of cosmological models.

  The model we shall use in this paper for the damped \lya systems
consists of two different parts. The first part specifies the mass
distribution of the dark matter halos where the damped absorbers are
assumed to originate, given a cosmological model for the primordial
density fluctuations; the second is the modeling of the gaseous
absorbers within each halo. For the first part, we adopt the
Press-Schechter formalism (Press \& Schechter 1974; Bond \etal 1991),
incorporating the usual cosmological parameters and the power spectrum
$P(k)$. We use the parameter $\delta_{c,t}=1.69$ (with a top-hat filter)
for the threshold overdensity required to form a halo (see, e.g., Gross
\etal 1997 for a recent investigation of the required value of
$\delta_{c,t}$ to fit the results of numerical simulations).
For the second part, our model for the gas distribution within a halo
is a spherically symmetric exponential profile for the mean gas density,
but with the gas distributed in discrete internal components. The
internal model will be completely specified after four quantities are
fixed: the core radius $r_g$ of the distribution of gas clouds,
the fraction of baryons in the halo $f_{HI}$
which are in gas clouds, the cloud covering factor per unit path 
length $c({\mathbf x})$ along a line of sight, and
the internal cloud velocity dispersion $\sigma_i$. These parameters 
will
be fixed by observational quantities as far as possible. Artificial
spectra through random lines of sight in halos will then be generated,
containing superposed absorption components, and they will be analyzed
in the same way as the observed spectra in PW to
test the model against the observed statistical properties of the
absorbers.

\subsection{Internal Gas-Cloud Model}

  For each halo of mass $M$ in the Press-Schechter formalism, we adopt
the dark matter density profile of Navarro, Frenk, \& White
(1996 \& 1997, hereafter NFW). The radius $r_{200}$ of the halo is
defined by $M = 200 \rho_{crit} (4/3) \pi r_{200}^3$,
and the circular velocity by $V_c^2=GM/r_{200}$
(the relation to $r_{vir}$ used above is $r_{vir}= 1.04 r_{200}$;
notice that $V_c^2 = GM/r_{vir}$ is often used, but the
difference is small). We use
the procedure of NFW to specify the internal parameters of this
profile, following the steps outlined in their Appendix.   

Our model for the collapsing gas consists of assuming that in each 
halo in the
Press-Schechter formalism with a given $V_c$, there is a 
spherically symmetric system of gas clouds that
are moving randomly and are responsible for the individual
absorption components observed in the metal-line profiles. PW
claimed that a system of clouds in a spherical halo
is not consistent with the metal absorption profiles, which are often
asymmetric, and that rotation in a flattened disk is needed.
We shall find that in fact, these absorption profiles do not provide
an unequivocal signature of a rotating system, in agreement with
Haehnelt \etal (1998), and that a spherical model is consistent with
the data.
The clouds are assumed to contain all the observed neutral hydrogen,
accounting for a fraction $f_{HI}$ of the total baryon mass of the 
halo, but occupying a small fraction of the volume in the halo.
They may be pressure-confined by a hot medium, as in Mo (1994)
and Mo \& Miralda-Escud\'e (1996),
or they may be gravitationally confined, or may originate in
satellites that are being disrupted (Morris \& van den Bergh 1994;
Wang 1993).
We choose the following model for the {\it mean}
density profile of the gas $\rho(r)$
(the internal gas density in an individual
cloud should of course be higher by the inverse of the volume
filling factor):
\begin{equation}
\rho(r) = \rho_o \exp(-r/r_g) ~,
\label{denseprof}
\end{equation}
where $r_g$ is a core radius. The parameters $r_g$ and $f_{HI}$ need
to be adjusted to reproduce the observed column density distribution
of damped \lya systems.

\subsection{Fitting the Observed Column Density Distribution}

\begin{figure}
\centerline {
\epsfxsize=4.7truein
\epsfbox[70 32 545 740]{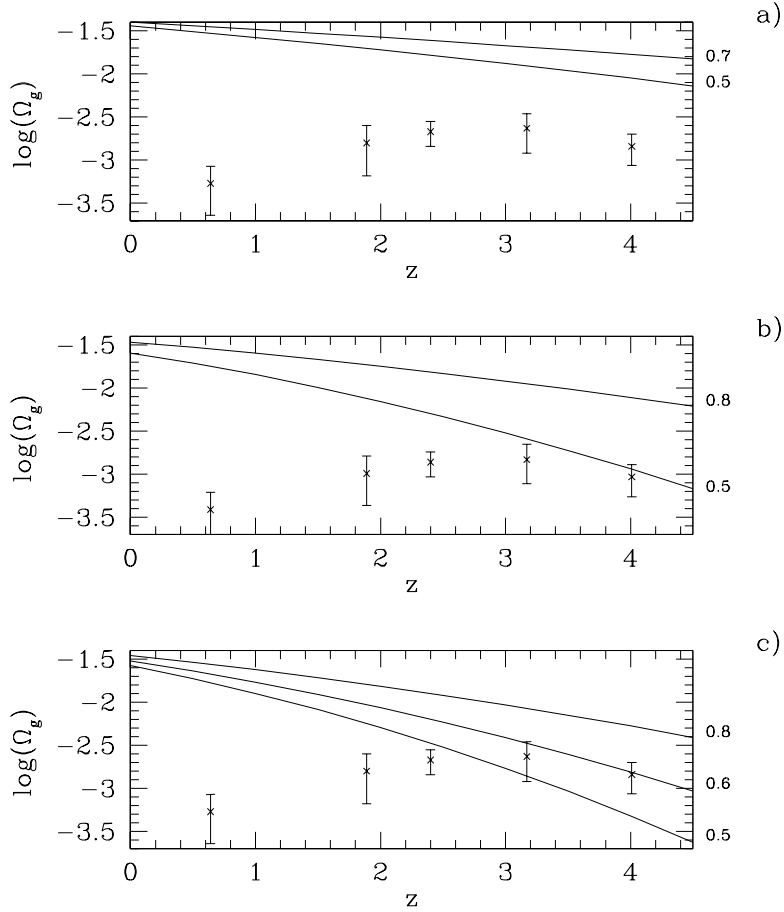}
}
\caption{Solid lines are the total baryons in collapsed objects
predicted in the following three models: (a) SCDM, (b) $\Lambda$CDM,
and (c) MDM, for
various normalizations of the power spectrum (the value of $\sigma_8$
at present is given at the end of each line on the right side; other
parameters of the models are given in Table 1), and
for $f_{HI}=1$.
Points with error bars are the observed $\Omega_g$
from Storrie-Lombardi \etal (1996). The circular velocity cutoff
is $V_{min} = 50 \kms$ in SCDM and $\Lambda$CDM, and $V_{min}=35 \kms$
in MDM.}
\label{OHI}
\end{figure}

\begin{figure}
\centerline {
\epsfxsize=4.8truein
\epsfbox[70 32 545 740]{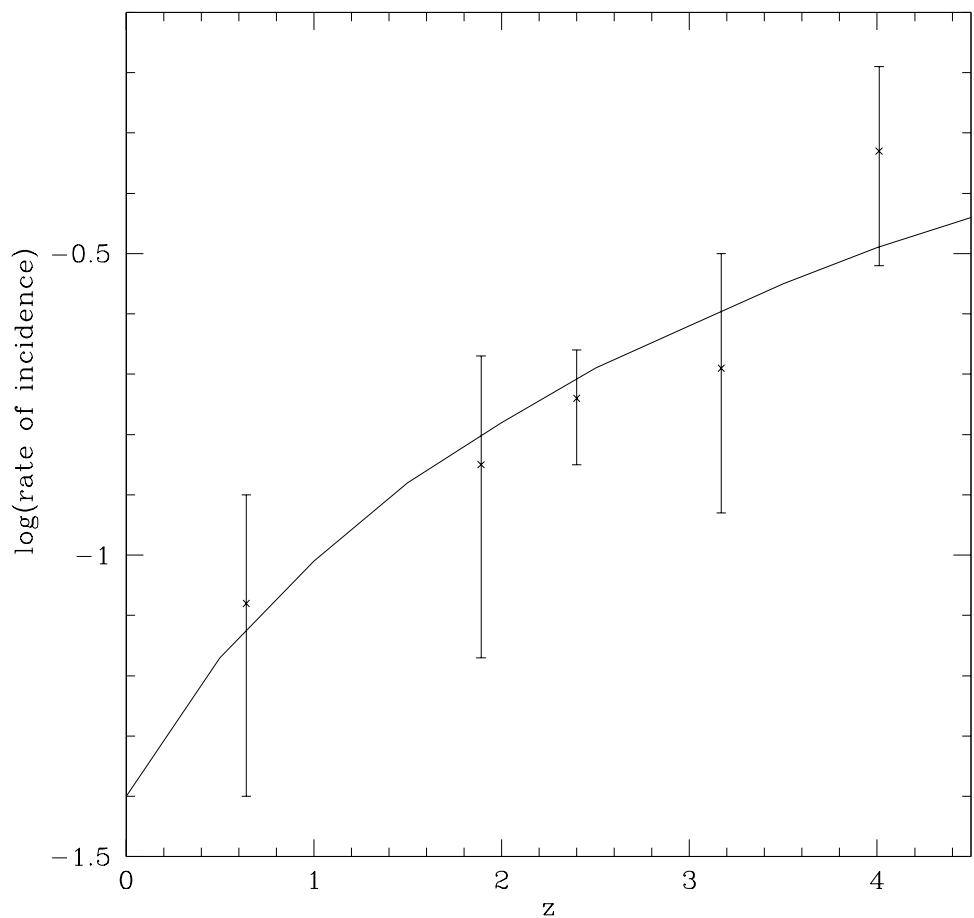}
}
\caption{Points with error bars are the observed incidence rate, or
number of systems with $N_{HI} > 2\times 10^{20}\cm^{-2}$ per unit
redshift, from Storrie-Lombardi \etal (1996). The solid line
is their power law fit to the data.}
\label{incrat}
\end{figure}

The parameters we fit directly are $c_g\equiv r_g/r_{200}$
and $f_{HI}$. We assume that, at a fixed redshift, the two parameters
are constant for all halos with different $V_c$, but they depend on
redshift. The observations to be matched are the fraction of the 
critical density in the form of atomic gas in damped \lya systems,
which we denote as $\Omega_g$, and the incidence rate of absorbers per
unit redshift. These are shown in Figures \ref{OHI} and \ref{incrat},
respectively, from the results of Storrie-Lombardi \etal (1996).

  To fit the two parameters at a given redshift, we first compute the
impact parameter in a halo where the column density is equal to the
threshold for an absorber to be considered a damped system
($N_{min}=2\times 10^{20} \cm^{-2}$), as a function of $V_c$, and then
compute the incidence rate and $\Omega_g$ by integrating over $V_c$
the mass of gas within this impact parameter
and the cross section of every halo.
The integration is carried out from a minimum value
$V_{\min}$ to infinity. It is expected that
in very small halos, with $V_c \lesssim 50 \kms$, the collapse of the
photoionized gas will be suppressed by the gas pressure, preventing
the formation of damped absorption systems (e.g., Efstathiou 1992;
Quinn, Katz, \& Efstathiou 1996; Thoul \& Weinberg 1996; Navarro \&
Steinmetz 1997). We normally use the cutoff $V_{min}=50\kms$, and find
the values of $f_{HI}$ and $c_g$ that give the observed
$\Omega_g$ and incidence rate for every possible normalization of
the power spectrum in each model.

  As the normalization of the power spectrum is lowered, the number
of massive halos in the model decreases until the fraction $f_{HI}$
reaches unity at some redshift. This usually happens first at the
highest redshift points in Figure \ref{OHI} and \ref{incrat}, and
it indicates that the model does not have enough power to explain
the observed amount of mass in damped systems. At this point, we
fix the parameter $f_{HI}=1$, and decrease $V_{min}$ to lower values
until the model prediction reaches the lower end of the $1-\sigma$
error bar in Figure \ref{OHI}. As we shall see in \S 4, as the
normalization is lowered the models fail the new kinematic tests
we present later before it is necessary to reduce $V_{min}$ (i.e,
before $f_{HI}$ reaches unity). Thus,
our constraints on the cosmological models including the observations
of the kinematics of the absorbers in PW are superior to the previous
ones, which were based only on the total amount of neutral hydrogen
contained in the damped \lya systems.
 
\subsection{Generation of the Velocity Profiles}

  Once the parameters $c_g$ and $f_{HI}$ are fixed at a given
redshift, we make random realizations of spectra of absorption
systems in the following way. The circular velocity of the halo, 
$V_c$,
is chosen from the Press-Schechter distribution, and a random
line of sight through the halo is selected. We then generate absorption
systems (``clouds'') along the line of sight, with probability per
unit length equal to $c({\mathbf x})$, and column density 
$N_{HI} = \rho({\mathbf x})/
c({\mathbf x})$. The quantity $c({\mathbf x})$ here is the covering 
factor, a new parameter
of the model that will be described in detail in \S 3. Once the
individual clouds are generated, we first check that the total column
density is larger than $N_{min} = 2\times 10^{20} \cm^{-2}$; if it is
not, the line of sight is rejected. Velocities of each cloud are then
generated from a Gaussian distribution, with dispersion equal to the
local velocity dispersion at the radius of the cloud, calculated from
Jeans' equations and assuming an isotropic velocity  dispersion
(a more exact procedure would be to construct the phase space density
of clouds depending only on the energy, but we adopt our more simple
model; notice that the assumption of a steady-state distribution is
also not exact, due to the on-going merging and dissipation). We assume
the clouds move in the potential well of the dark matter halo plus
their own mass density, although in practice the contribution of the 
gas to the
gravitational potential is negligible in the models that satisfy the
observational requirements. The dark matter profile is that of NFW,
which is a good fit to the halo profiles obtained from numerical
simulations. Each cloud in a realization is assumed to produce an
absorption component with a Gaussian velocity distribution 
(due to internal thermal
or turbulent velocities), with velocity dispersion
$\sigma_i = 4.3 \kms$, as in PW. This value of $\sigma_i$ gives a
good fit to the typical widths of the absorption components.
We have verified that the full Monte Carlo procedure produces a 
column density distribution that closely matches the observed one.

\section{The Dissipation Rate in Damped \lya Systems}

The velocity structure of the damped \lya systems, as revealed in
associated metal absorption lines, can generally be modeled as the
superposition of one or a few absorption components. If these components
are due to clouds moving in a halo, this implies that the clouds should
undergo collisions with each other several times in every orbit around
the halo, because along the path of the motion of an individual cloud
within the halo several clouds should be intersected, just as is the
case for a random line of sight.
\footnote{In a rotating system of clouds, some of the velocity
dispersion along a line of sight may be contributed by a variation of
the rotational velocity within the system rather than a local dispersion
in velocities, but as we shall see later the contribution of rotation is
generally not much larger than that of the local dispersion.}
In every such collision,
a large fraction of the kinetic energy in the center-of-mass frame of
the colliding clouds is converted to
thermal energy in a shock and then radiated.
The multiple absorbing components often have similar contributions
to the total column density of the system, and therefore a large fraction
of the total kinetic energy of the clouds must be dissipated every
orbital time.
Since the damped \lya
systems also contain a large fraction of the collapsed baryons in the
universe, we conclude that
{\it the amount of energy being dissipated in the
absorption components of damped \lya systems is necessarily very large}.
The smaller the absorption regions are, the faster the dissipation rate
because the orbital times are shorter.

  It needs to be said here that the ``clouds'' in our model are only
supposed to be a concept that helps to define a simple approximate
interpretation of the observations. In reality, the neutral gas may well
have a continuous density distribution, and the cloud collision rate in
our model would then represent the frequency of strong shocks in the
gas, where the thermal energy produced should be rapidly radiated 
away.

  This large energy dissipation rate in damped \lya systems is our
motivation to propose the following hypothesis: damped \lya systems 
are the site where most of the negative gravitational potential 
energy of the baryons in galaxies was dissipated when galaxies 
formed, and the gas in these systems is in the process of dissipating 
this energy. Given this assumption, we develop in this section a 
method to predict the covering factor of the absorption components.
This hypothesis will be further discussed at the end of the paper 
in light of the results to be presented in \S4.
In the rest of this section, we calculate the rate of energy release 
in a halo due to mergers and the rate at which energy is dissipated in
cloud collisions.  The two rates must be equated under the proposed
hypothesis.

\subsection{Rate of energy generation by gravitational collapse.}

In the absence of dissipation, gas should
be shock-heated to the halo virial temperature and should settle
to a density profile similar to that of the dark matter (e.g., Navarro
\etal 1995).
Dissipation causes the gas to become more concentrated
than the dark matter. The gas profile in our model is fixed to the
form in \eq  (1), and is indeed required to be more concentrated than
the dark matter to fit the observed neutral hydrogen column density
distribution.
It is straightforward to calculate the difference in
energy between the two configurations of the gas. 
Initially, the gas and dark matter are assumed to be in the same (NFW)
profile and the total energy of the system is $E_1$.  A 
fraction $f_{HI}$ of the gas
then collapses into the profile in \eq  (1). 
The dark matter distribution also contracts as a result, as required by
adiabatic invariance.  The resulting system has energy
$E_2$. The difference between initial and collapsed energies must be
the total energy that has been dissipated in this halo:
\begin{equation}
E_{d} = -(E_2-E_1) ~.
\end{equation}
The energies are calculated using
\begin{equation}
E= -\frac{G}{4}
{\, \frac{M^2}{r_{200}}+
\int_0^{r_{200}}\left[{M(r)\over r}\right]^2 dr\, } ~,
\label{enint}
\end{equation}
with $M(r)$ the total mass (gas plus dark matter) inside radius $r$.
For the NFW profile this yields
\begin{equation}
E_1=-\frac{G M^2}{2 r_{200}} f_c ~,
\end{equation}
with 
\begin{equation}
f_c=\frac{c}{2}\left[\frac{1-1/(1+c)^2-
2 \ln (1+c)/(1+c)}{c/(1+c)-
\ln (1+c)}\right] ~,
\end{equation}
where $c=r_{200}/r_s$ is the concentration parameter of the 
NFW profile (Mo, Mao, \& White 1998).  This parameter depends on the 
formation time of the halo as described in Navarro \etal (1997).
The dark matter density as a function of radius is 
\begin{equation}
\rho_{NFW}=\frac{\rhoc \delta_c}{r/r_s(1+r/r_s)^2} ~,
\end{equation}
with
\begin{equation}
\delta_c=\frac{200}{3}\frac{c^3}{\left[\ln\left(1+c\right)-
         c/\left(1+c\right)\right]} ~.
\end{equation}	 

The energy after collapse is calculated using the assumed exponential
gas profile, and the dark matter profile is calculated using
adiabatic invariance,
\begin{equation}
M(r)\, r=M(r_i)\, r_i ~.
\end{equation}
Here, $M(r)$ is the total mass inside the shell
of radius $r$ after collapse, $r_i$ is the initial radius of the dark
matter shell which has contracted to radius $r$, and $M(r_i)$ is the
total mass initially within this radius (when the gas and dark matter
had the same NFW profile). With the additional equation (required by
mass conservation)
\begin{equation}
M(r)= M(r_i)\cdot(1- f_b f_{HI}) + M_g(r) ~,
\label{masscon}
\end{equation}
where $M_g(r)$ is the gas mass inside radius $r$, we
can solve for $r_i$, $M(r_i)$ and $M(r)$ for a given radius $r$.
Finally, the energy can be found by integrating equation 
(\ref{enint}).
Notice that we have assumed here that only a fraction $(1-f_b f_{HI})$
of the mass undergoes the collapse to the profile of the observed
atomic gas. The rest of the baryons (a fraction $1-f_{HI}$) might be
in stars or dense molecular clouds, and may undergo its own loss
of energy during mergers. The interaction of this other baryonic
component with the atomic gas can be very complex: the dense molecular
gas may transmit some energy to less dense gas through shocks. 
In addition, the gravitational energy of galaxies in merging halos can
generally be given off to less dense dark matter and gas through
dynamical friction. 
Here, we need to neglect these processes for simplicity. Numerical
simulations should be useful in investigating the possible
importance of these effects.

We now use an extension of the PS model (Bower 1991; Lacey \& Cole 1993)
to compute the rate of energy production from gravitational collapse in
each halo. Lacey \& Cole (1993) give the average rate of merging of a
halo of mass $M_0$ with a halo of mass $\Delta M$ to form a halo of mass 
$M=M_0+\Delta M$: 
\begin{equation}
\frac{d^2 p}{d\ln\Delta M dt}(M_0, M) = 
\sqrt{\frac{2}{\pi}} \,
\left|\frac{d\delta(t)}{dt}\right|\, \frac{\Delta M}{\sigma_{M}^2} \,
\left|\frac{d\sigma_{M}}{dM}\right| \,
\frac{1}{\left(1-\sigma_{M}^2/\sigma_{M_0}^2\right)
^{3/2}} \,
\exp\left[-\frac{\delta(t)^2}{2}\left(\frac{1}{\sigma_{M}^2}-
\frac{1}{\sigma_{M_0}^2}\right)\right] ~,
\end{equation}
where $\delta(t)=1.69/D(t)$ is the threshold for collapse, 
$(\sigma_M, \sigma_{M_0})$ are the present $rms$ 
overdensity in linear theory inside a sphere containing mean
mass $(M, M_0)$, and $D(t)$ is the linear growth factor
normalized to unity at the present time.
Fits for $\sigma_M$ are taken from 
Ma (1996) and Navarro \etal (1997).
The comoving number density of halos of mass $M$ is given by
\begin{equation}
\frac{dn}{dM}=\sqrt{\frac{2}{\pi}}\, \frac{\rho_0}{M} \,
\frac{\delta(t)}{\sigma_M^2}\, \left|\frac{d\sigma_M}{dM}\right| \,
\exp\left[-\frac{\delta(t)^2} {2 \sigma_M^2} \right] ~,
\end{equation}
where $\rho_0$ is the mean density of the universe.

The mean rate of energy delivered to a halo of mass $M$ by
gravitational collapse can be written as
\begin{equation}
M\frac{d(E_d/M)}{dt}=\left.\frac{\partial E_d}{\partial t}\right|_M+
\int_{M/2}^{M} \left\{ \left[ E_d(M) - E_d(M_0) - E_d(\Delta M) \right] \,
\frac{d^2 p}{d\Delta M dt}(M_0)
\frac{dn}{dM}(M_0) dM_0\right\} \, \left[ \frac{dn}{dM}(M) \right]^{-1} ~,
\label{rateofgen}
\end{equation} 
where $\Delta M=M-M_0$.
The first term in equation (\ref{rateofgen}) is the average rate of
variation of the energy for a halo of constant mass, and
the second gives the mean increase in the energy due to the
probability that the halo of mass $M$ merged from halos of mass $M_0$ and
$\Delta M$.
The first term includes the change with time of the fit parameter
$c_g(z)$, the concentration parameter of
the NFW profile $c(z)$, and $r_{200}$, all for a fixed mass $M$.  
In practice, the second term in equation (\ref{rateofgen}) is more
important than the first, owing to the rapid increase of the velocity
dispersion of collapsed halos at high redshift.

\subsection{Rate of energy dissipation by the gas.}

  We now calculate the energy dissipation rate due to cloud collisions.
Assuming perfectly inelastic collisions of clouds of equal mass $m$,
moving at velocities ${\mathbf v}_1$ and ${\mathbf v}_2$,
the energy lost in a collision is
\begin{equation}
\Delta E = \frac{m}{4}|{\mathbf v}_1-{\mathbf v}_2|^2 ~.
\end{equation}
We assume the velocity distribution function for the clouds to be
\begin{equation}
f({\mathbf x},{\mathbf v})= \frac{n({\mathbf x})}{[\sqrt{2 \pi}
               \sigma({\mathbf x})]^3} 
                e^{-v^2/(2 \sigma^2({\mathbf x}))} ~,
\end{equation}
where $\sigma({\mathbf x})$ is the velocity dispersion at any point
${\mathbf x}$ in the halo. If $s({\mathbf x})$ is the cross section for
cloud collisions, and $n({\mathbf x}) = \rho({\mathbf x})/m$ is the cloud
number density, the rate of energy loss per unit volume in cloud
collisions at point ${\mathbf x}$ is
\begin{eqnarray}
P({\mathbf x})= s({\mathbf x})\int 
d^3{\mathbf v} d^3{\mathbf v}'
f(v) f(v') \,\frac{m}{4}|{\mathbf v}-{\mathbf v'}|^2\,
|{\mathbf v}-{\mathbf v}'|  \\ \nonumber =
\frac{8}{\sqrt{\pi}}\, \rho({\mathbf x})\, s({\mathbf x})\,
n({\mathbf x})\, \sigma^3({\mathbf x}) ~,
\end{eqnarray}
and the total rate of energy dissipation in the halo is
\begin{equation}
\frac{dE}{dt}=\int P({\mathbf x})d^3{\mathbf x} ~.
\end{equation}
We choose the cross section so that the mean time between 
collisions $\tau_c$ for any given cloud is constant throughout the
halo, i.e., $n({\mathbf x})\, s({\mathbf x})\, \sigma({\mathbf x}) =
{\rm constant} \equiv \tau_c^{-1}$.
The rate of energy dissipation can then be rewritten as
\begin{equation}
\frac{dE}{dt}= \frac{8}{\sqrt{\pi}}\, \tau_c^{-1}
\int \rho({\mathbf x})\, \sigma^2({\mathbf x})\, d^3{\mathbf x} ~.
\label{endisrate}
\end{equation}
This can be compared to the rate of energy generation computed in 
\S3.1. The model also specifies the covering factor of the absorbers,
$c({\mathbf x})=[\tau_c\, \sigma({\mathbf x})]^{-1}$.
Random realizations of the absorption spectra are computed as
described in \S 2.3, adding absorbers along a random line of sight
with probability per unit length $c(\mathbf x)$ and column density
$N_{HI} = \rho({\mathbf x})\, \sigma({\mathbf x}) \, \tau_c$.

\section{Results}

In the previous two sections we have presented a generic spherical
halo model for absorbing clouds in damped \lya systems. The spherical
model completely specifies the properties of the absorbing clouds in a
given halo, as a function of four parameters:
the fraction of baryons in neutral gas $f_{HI}$, the radius of the
cloud distribution $r_g$, the internal cloud velocity dispersion
$\sigma_i$ (which will be fixed to $\sigma_i=4.3\kms$ throughout this
paper), and the covering factor of clouds. 
For any large-scale
structure theory specifying the number of halos with circular
velocity $V_c$ at every redshift, we can calculate any 
desired property of the absorption profiles by 
fitting the values of $f_{HI}$ and $r_g$ to the observed
mass and incidence rate of the damped \lya systems, for different 
values of the parameter $\tau_c$ in \S3.2 determining the covering factor
of the clouds.  We can then see if the value of $\tau_c$ derived from the
energy balance results in a cloud covering factor consistent with 
observation.

\subsection{Fit Values for $f_{HI}(z)$ and $c_g(z)$}

Our first step is to fit for $f_{HI}(z)$ and 
$c_g(z)\equiv r_g/r_{200}$.
Figure \ref{OHI} shows the observed 
$\Omega_{g}$ in three models  
(the differences are the result of the different
pathlengths for a fixed redshift range in different
models), and Figure 2 shows the observed incidence rate we use, from
Storrie-Lombardi \etal (1996) (the solid curve is their power-law fit
to the data).
The cosmological models we use are defined in Table 1; all the models
have $\Omega_b=0.05$, $h=0.65$, and $n=1$. We have taken fitting
formulae for the power spectra from Ma (1996) and Navarro \etal (1997).
The normalization of the power spectra (parameterized as usual by
$\sigma_8$) will be varied for all models.
\begin{table}
\begin{tabular}{lcccccccc}
\tableline\tableline
Model & $\Omega_0$ & $\Omega_{\Lambda}$ & $\Omega_{\nu}$ & 
$\sigma_8$ &
$f_{HI}(z=2)$ & $f_{HI}(z=4)$ & $c_g(z=2)$ & $c_g(z=4)$  \\ \tableline
SCDM & 1.0 & 0.0 & 0.0 & 0.5 & 0.10 & 0.28 & 0.031 & 0.069  \\ 
$\Lambda$CDM & 0.4 & 0.6 & 0.0 & 0.77 & 0.076 & 0.23 & 0.034 & 0.096  \\ 
MDM & 1.0 & 0.0 & 0.2 & 0.8 & 0.15 & 0.65 & 0.036 & 0.12 \\ 
OCDM & 0.3 & 0.0 & 0.0 & 0.85 & 0.076 & 0.13 & 0.038 & 0.089  \\ \tableline
\end{tabular}
\label{models}
\caption{Model definitions and examples of gas profile parameters.}
\end{table}

\begin{figure}
\centerline {
\epsfxsize=4.8truein
\epsfbox[70 32 545 740]{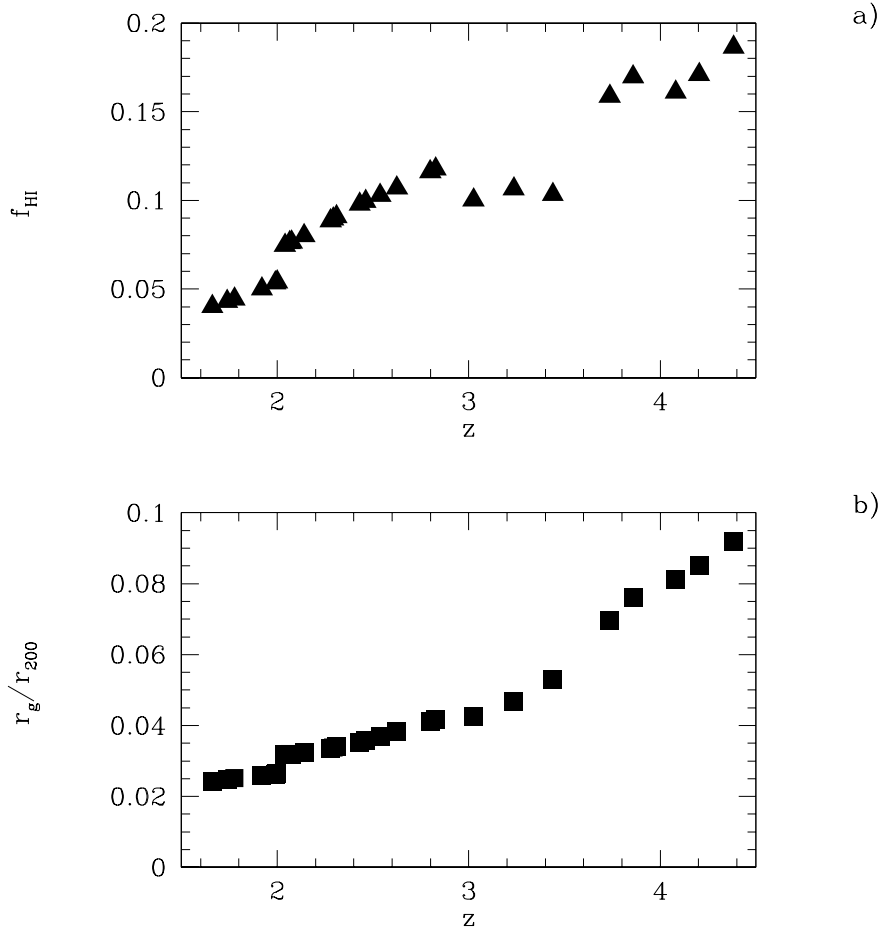}
}
\caption{Results of fitting our model parameters $f_{HI}$ and
$c_g\equiv r_g/r_{200}$ to the observations for 
a $\Lambda$CDM model with $\sigma_8=0.9$.  The discontinuities
in the fit values are the result of the binning in the $\Omega_g$
data that was fit.}
\label{fHIrgorh}
\end{figure}

The values of $f_{HI}$ 
and $c_g$ at different redshifts are shown in Figure \ref{fHIrgorh}
for the case of the $\Lambda$CDM model with $\sigma_8=0.9$. 
We also give the values of $f_{HI}$ 
and $c_g$ at $z=2$ and $z=4$ for all the models in Table 1. 
The normalization is the one required to fit the present cluster
abundances according to Eke \etal (1996) and Ma (1996).
In general, $f_{HI}$ and $c_g$ grow with redshift,
in agreement with the reasonable expectation that at earlier epochs
a greater fraction of the baryons should be in atomic gas, with a
more extended distribution in the halos.

\subsection{Energy Generated Through Gravitational Collapse}

The results in this Section are generally computed from randomly
generated absorption profiles of the low-ionization metal lines in
damped \lya systems, according to the model of clouds in a spherical
halo described previously. The spectra of absorption systems are
generated in groups of 27, at the same redshifts as the 27 absorbers in
``Sample C" of PW2, in order to compare the model predictions with the
observations of PW2. Values of $\Omega_g$ and the incidence rate are
interpolated in redshift from the points in Figures \ref{OHI} and
\ref{incrat}, to obtain $c_g$ and $f_{HI}$ at each redshift. Gaussian
distributed noise is added to each absorption system with an amplitude
corresponding to the signal-to-noise for this redshift system (given by
PW). Finally, following PW, the absorption profiles are smoothed by a
top-hat window 9 bins wide. We show later in Figure \ref{linesEE} 
an example of
27 random absorbers for a model that agrees with the observations.

\begin{figure}
\centerline {
\epsfxsize=4.8truein
\epsfbox[70 32 545 740]{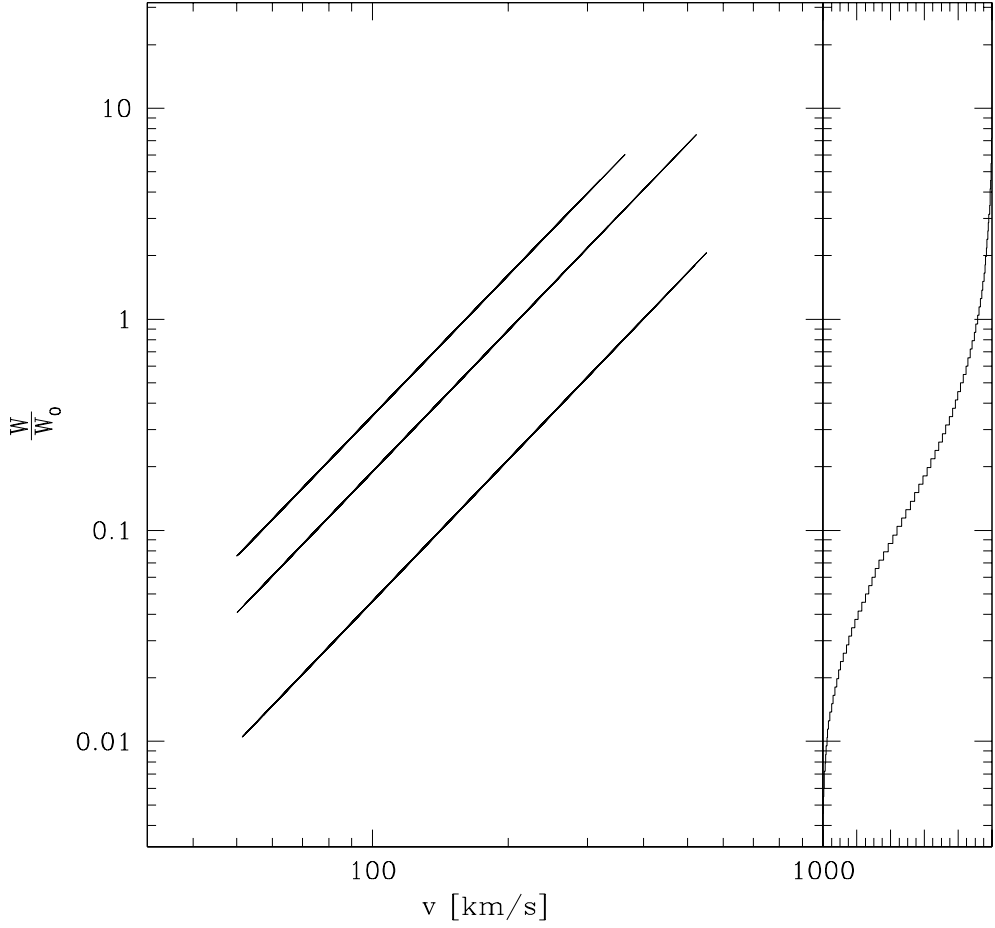}
}
\caption{For the $\Lambda$CDM model ($\sigma_8=0.8$), the rate of
energy production from gravitational collapse that we compute for 
halos at redshifts $z=2$, $z=3$, and $z=4$ (from bottom to top).
$W_0 \equiv 1\times 10^{50}
\erg(10^{12}\msun)^{-1}\,{\rm yr}^{-1}$.
The panel on the right is the cumulative number of systems from
a randomly generated set of 200 halos at each of the 27 observed
redshifts. }
\label{fenergies}
\end{figure}

Figure \ref{fenergies} is an example of the rate of energy production
per unit mass from gravitational collapse as a function of halo size 
$V_c$ for the $\Lambda$CDM model. The rate $W\equiv d(E_d/M)/dt$
is computed using equation (\ref{rateofgen}). We define the unit
$W_0=10^{50}\erg\,{\rm yr}^{-1}/(10^{12}\msun)$ to facilitate
comparisons between this rate of energy generation and other possible
sources; for example, if a supernova explosion injects $10^{51}$ergs
of energy, $W_0$ corresponds to one supernova per ten years in a
halo of mass $10^{12}\msun$.

  We now make a test of our hypothesis that the energy dissipated in
cloud collisions in damped \lya systems is provided by gravitational
accretion, as explained in \S 3. Equating the energy generation rates
in equations (\ref{rateofgen}) and (\ref{endisrate}), we obtain the
parameter $\tau_c$ which determines the covering factor, or the typical
number of components in the randomly generated absorbers. Following PW1,
for each of the simulated absorption profiles we identify the point of
maximum optical depth, $\tau_1$, as the first peak. If the optical depth
has other maxima, we say there is a second peak if the next highest
maximum has optical depth $\tau_2 > \tau_1/3$ and if $\tau_2$ is at
least a $3\sigma$ feature above the minimum optical depth in the
velocity interval between the two peaks. Here $\sigma$ is the noise
level of each of the 27 observed spectra.

  Of the 27 systems in PW2, 10 had a second peak. We have randomly 
generated 200 sets of 27 absorption profiles for every model in 
Table 1.  We find that the mean number of systems (out of 27) in each
set with a second peak is typically $\sim 1$.
This means that
{\it the energy computed from gravitational collapse}
(Fig. \ref{fenergies}) {\it is not sufficient to support the
multicomponent structure of the observed velocity profiles}. 
When we match the rates of energy
generation from gravitational collapse with the rates of energy 
dissipation by cloud collisions, we find that
the line profiles are almost exclusively single Gaussians with 
$\Delta v \sim 20\kms$. The amount of energy provided by 
gravitational collapse is far too small to provide the energy dissipated
in a cloud collision at every orbital time, so most lines of sight
intersect only one cloud. This result is independent of the cosmological
model we adopt. Different cosmological models cause only slight
modifications of the rate of energy injection per unit mass $W$.

A point that needs to be made here is that the model of
a rotating disk may not greatly alleviate the requirement of a fast
dissipation rate of the kinetic energy contained in the absorbing
clouds. This requirement was not considered in the disk models of PW,
because their models were not dynamically self-consistent: the disks of
clouds were required to be thick and cold at the same time, so that the
relative velocities of clouds on a random line of sight were dominated
by the variation of the rotational velocity along the line of sight.
In a dynamically self-consistent disk, the clouds must have the vertical
velocity dispersion that is determined by the vertical gravity in the
disk. In the limiting case where the vertical component of gravity is
determined by a spherically symmetric dark matter halo and the disk
mass is negligible, the vertical velocity dispersion is
$\sigma_z = V_c\, h/(2^{1/2}R)$, where $h$ is the scale-height and $R$
the radius in the disk. The radial and azimuthal velocity dispersions
should generally not be smaller than $\sigma_z$. The dispersion in the
projected component of the circular velocity along the line of sight,
within the interval where the line of sight is contained within a
height $h$, is $V_c\, h/(2R\, \tan i)$, where $i$ is the inclination of
the line of sight from the disk plane (we have assumed here a constant
$V_c$ with radius as observed in most disk galaxies, but the answer is
not very sensitive to reasonable changes in the rotation curve).
If the angle $i$ is not very
small, rotation cannot decrease the local velocity dispersion of the
clouds compared to the observed line of sight velocity dispersion by a
large factor. Moreover, any effects of rotation on the morphology of the
absorption profile will necessarily have to be washed out to a
significant degree by the internal cloud velocity dispersion. If
self-gravity of the disk is important, $\sigma_z$ increases, making
the cloud velocity dispersion more important than rotation. A
possibility that needs to be examined more carefully is that most cases
of multiple absorbing components in damped \lya systems correspond to
lines of sight intersecting a disk with $i\ll 1$; excepting this
possibility, rotational velocities cannot predominate over the internal
cloud velocity dispersion in yielding the observed line of sight
velocities.

\subsection{Energy Injection from Other Sources}

In general, sources of energy other than the gravitational collapse
of structure may exist to maintain the random motion of the clouds in
damped \lya systems, such as the energy from supernova explosions when
stars form in a galaxy. Simple estimates (e.g., Bookbinder \etal 1980)
show that the total energy liberated in supernova explosions
when a large fraction of the gas turns to stars is sufficient to heat
the remaining gas to temperatures $\sim 10^7$ K. Even if a large 
fraction of the supernova energy is lost in radiation, it is clear 
that if a wind can be blown into the halo after episodic starbursts,
supernovae could be important in replenishing the kinetic energy in the
motion of the halo clouds, since the typical relative velocities among
the absorption components are $\sim 100 \kms$, corresponding to virial 
temperatures of only $10^6$ K. 
A greater supply of energy can support more cloud collisions, and 
thus a more frothy, high covering factor structure for the density
distribution.
This higher covering factor leads to a more complete sampling of the 
underlying velocity distribution, 
which produces larger velocity widths in the absorbers.  

\begin{figure}
\centerline {
\epsfxsize=4.8truein
\epsfbox[70 32 545 740]{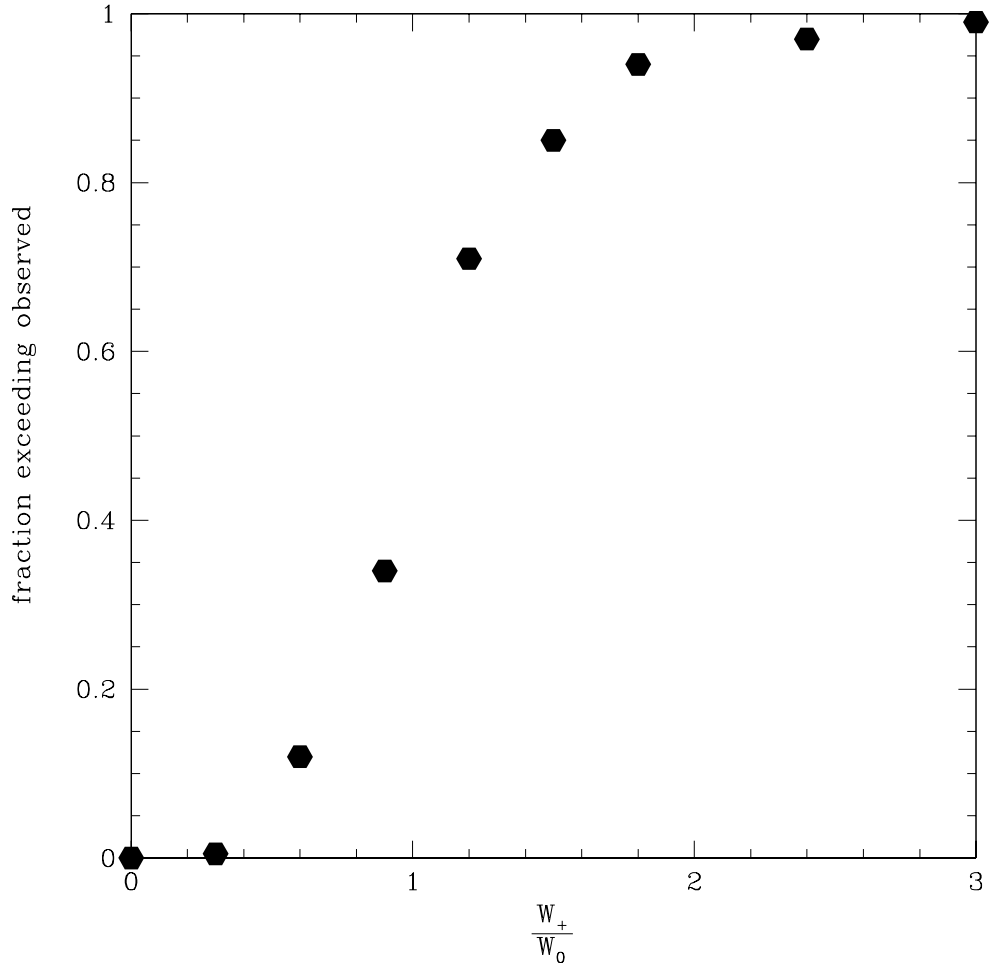}
}
\caption{Fraction of simulated sets of 27 systems which contain
more than the observed 10 lines with second peaks. $W_+$
is the rate of extra energy injection per mass (beyond the energy 
from gravitational collapse). 
The model is SCDM, $\sigma_8=0.7$.}
\label{covfac}
\end{figure}

We can parameterize this uncertainty 
by defining a rate of energy injection per unit mass, $W_+$, which we 
add to the rate of energy generation from gravitational merging 
in equation (\ref{rateofgen}).
Notice that, by adding a constant $W_+$ to all
halos, we increase the rate of energy injection by a much larger factor
in small halos, because small halos have a lower gravitational energy
per unit mass due to their small velocity dispersion, and they also
accrete much less matter than large halos.
The results for the fraction of simulated sets of 27 systems with more
than 10 systems with a second peak (for the SCDM model)
are shown in Figure \ref{covfac}, as a function of $W_+$.
We see that the energy injection rate required for consistency with the
observed number of components is of order $W_0$. 

\begin{figure}
\centerline {
\epsfxsize=4.8truein
\epsfbox[70 32 545 740]{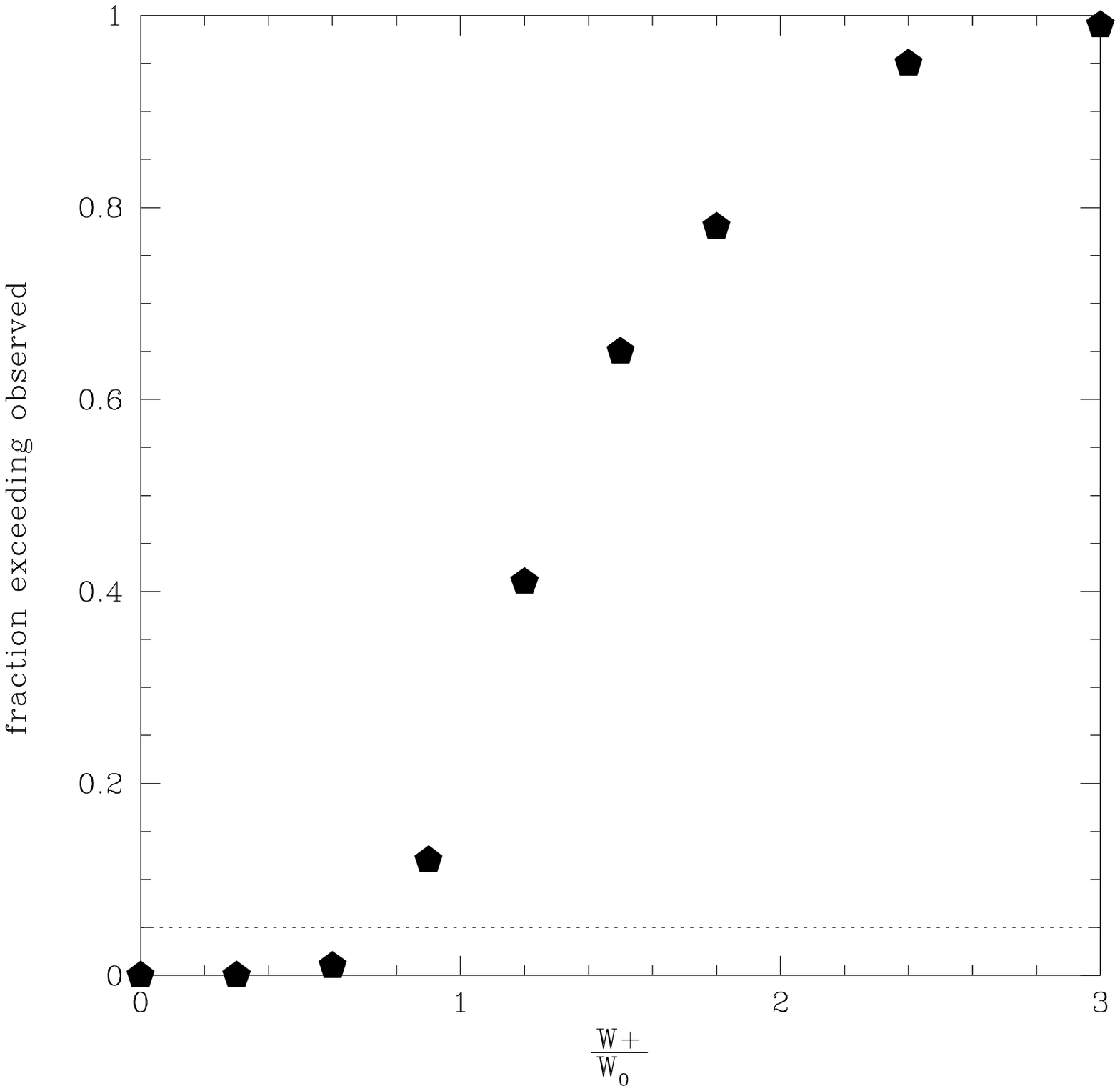}
}
\caption{For the example SCDM model with $\sigma_8=0.7$, the 
fraction of realizations
of the 27 observed systems which have median velocity width exceeding
the observed 77km/s.  
$W_+$ is the rate of energy injection per mass.}
\label{SCDMoverE}
\end{figure}

\begin{figure}
\centerline {
\epsfxsize=4.8truein
\epsfbox[70 32 545 740]{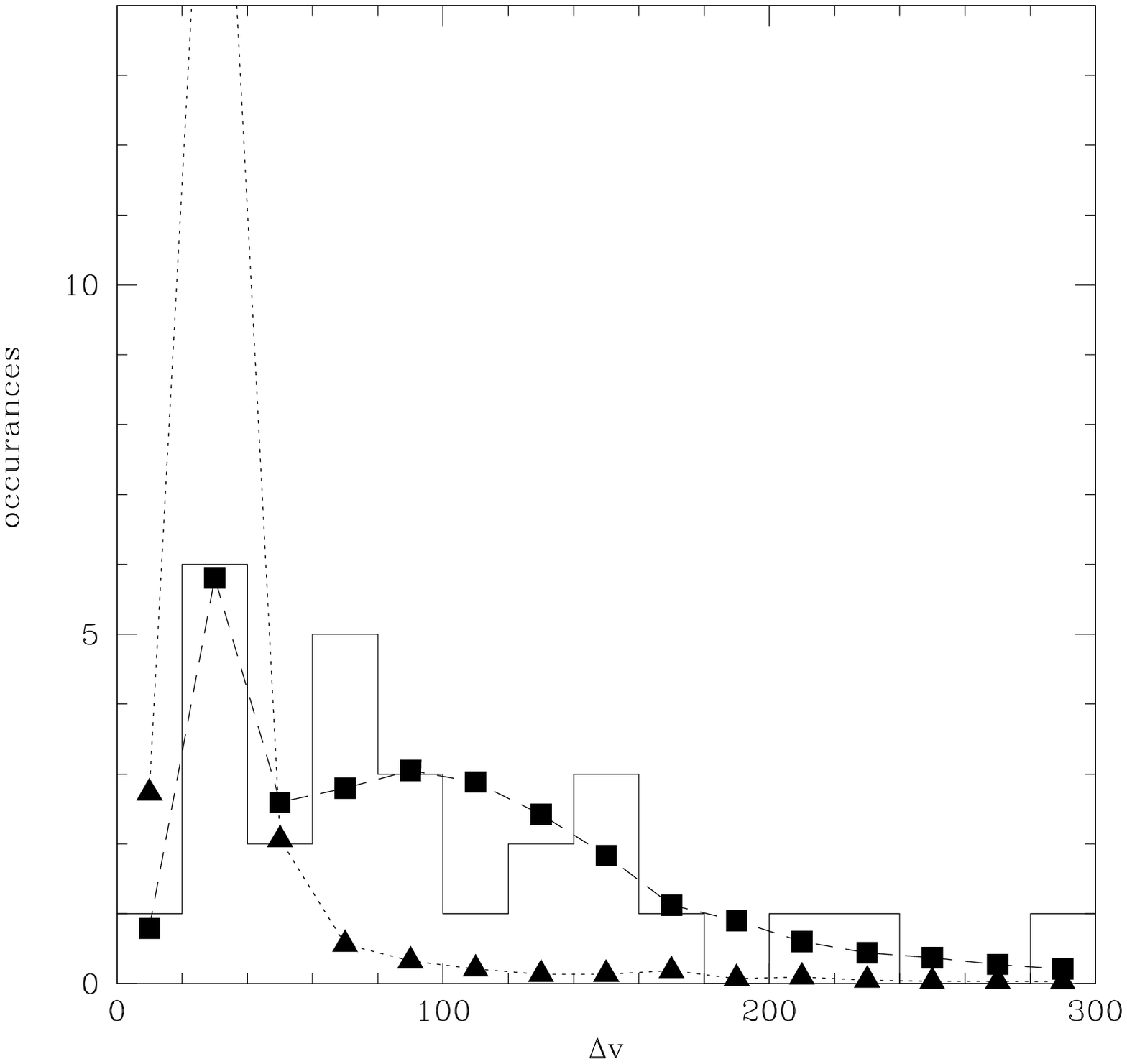}
}
\caption{Distribution of velocity widths. The 27 observed widths are 
shown by the solid histogram, the triangles show the predicted 
distribution for SCDM ($\sigma_8=0.7$)
with no energy injection, and the squares show the distribution for 
SCDM with energy injection rate $1.8 W_0$.}
\label{exvel}
 \end{figure}

\begin{figure}
\centerline {
\epsfxsize=4.8truein
\epsfbox[70 32 545 740]{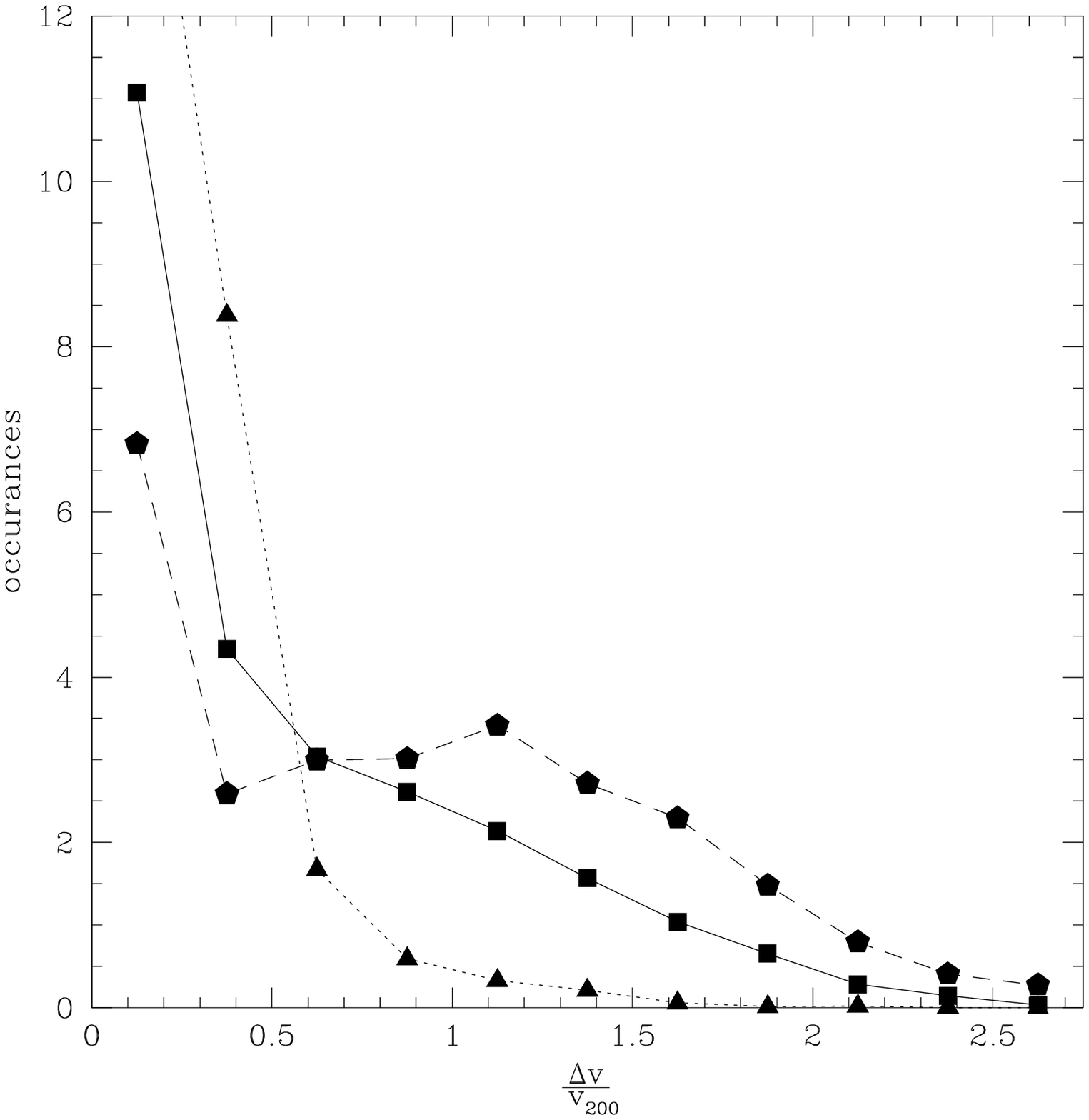}
}
\caption{Predicted distribution of the ratio of the velocity width to
the dark matter halo virial velocity. Using the SCDM ($\sigma_8=0.7$)
model, triangles, squares, and pentagons show energy injection rate
$W_+=0$, $W_+=0.6W_0$ and $W_+=1.8W_0$, respectively.}
\label{exvelovc}
 \end{figure}

The effect of increasing the number of components 
on the observed velocity widths [defined in equation (\ref{dvw})]
is shown in Figure \ref{SCDMoverE} (for the SCDM model).
The velocity widths are also consistent with observations for the same
range of the
rate of energy injection $W_+$. All the models require the extra energy
injection to agree with the observed velocity widths.
Figures \ref{exvel} and \ref{exvelovc} show in more detail the effect
on the distribution of velocity widths of
the rate of energy injection in our model. In most of the models that
have the right amplitude of density fluctuations on cluster scales to
fit observations of large-scale structure, the halos where
damped \lya systems arise are generally massive enough to explain the
observed velocity widths, given their velocity dispersions.
What is more difficult is to understand the source of the energy that
keeps the high level of turbulence of the gas.

  The implication of the result that energy injection is necessary will
be discussed further below.  For now we will assume that there is some
arbitrary amount of extra energy available and compare our model to the
observations using this assumption. Unfortunately this makes our model
less constraining because it introduces another free parameter.

\subsection{Overview of the Simulated Absorption Spectra}

\begin{figure}
\centerline {
\epsfxsize=4.8truein
\epsfbox[70 32 545 740]{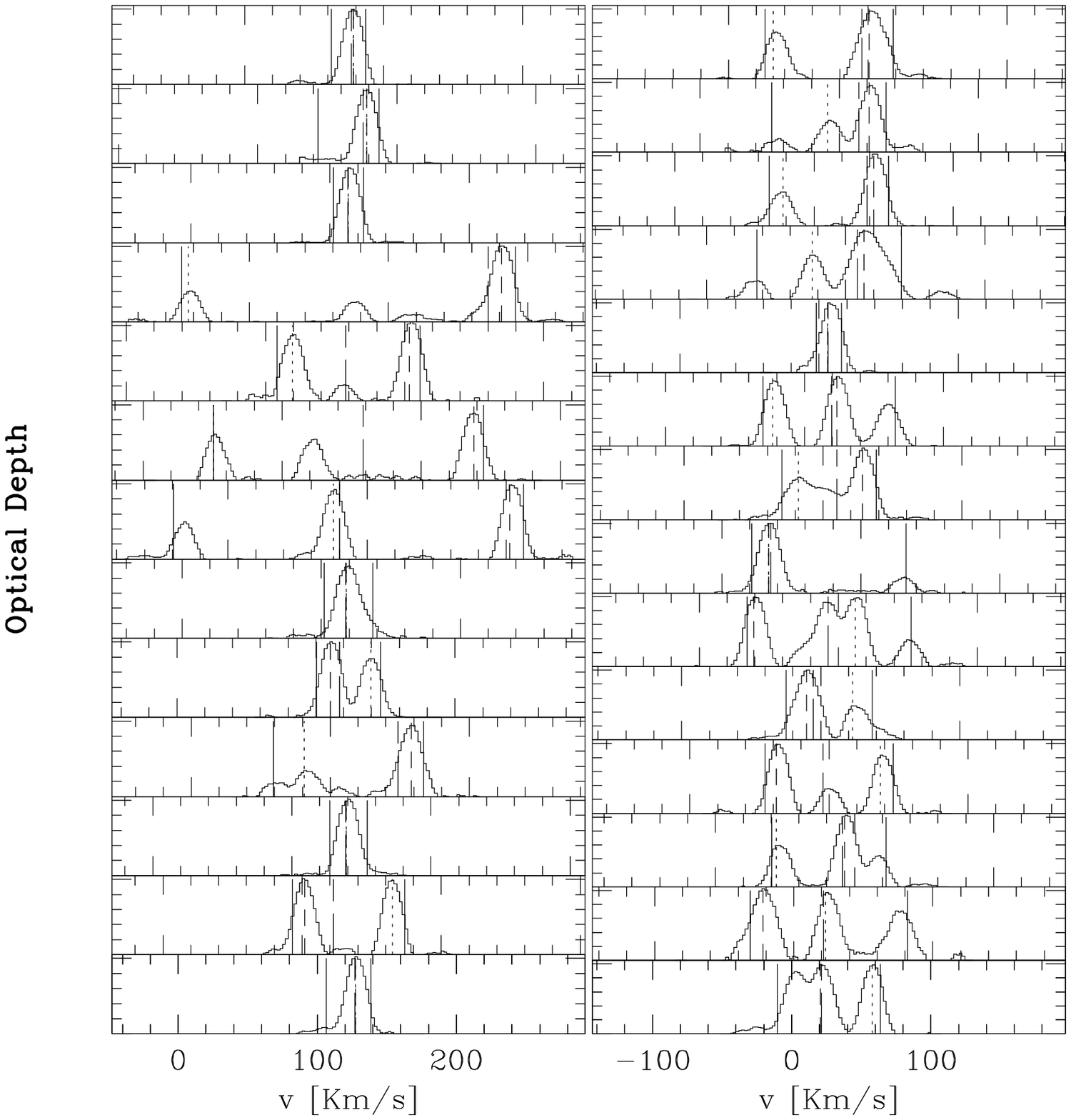}
}
\caption{
Simulated set of 27 absorption profiles 
($\Lambda$CDM, $\sigma_8=1.0$, energy injection rate $1.8W_0$).  
Solid lines exclude 5\% of 
the integrated optical depth
on each side.  Large dashed lines mark the median velocity, 
with 50\% of the integrated optical depth on each side.  
Medium dashed lines mark the maximum optical depth, and
the small dashed lines mark the second peak as defined in the text.
} 
\label{linesEE}
\end{figure}

  An example of a set of 27 absorption spectra is shown in
Figure \ref{linesEE}, for the $\Lambda$CDM model in Table 1 with
normalization $\sigma_8=1.0$, and energy injection rate $1.8W_0$.
This figure can be compared directly
with observed spectra in PW. Only the optical depth is shown here,
with arbitrary normalization.  Without energy injection, the
spectra do not resemble the observed ones at all because they 
are dominated by single Gaussian profiles.
But the spectra with energy injection bear remarkable
similarity to the observed profiles, as far as one can tell by simple
eyeball examination. 
Notice that our model is extremely
simplified; for example, we have assumed that every individual 
component has a Gaussian profile with a fixed width. 
In spite of this, the match to the observations is very good. 
We should bear in mind that two parameters have been adjusted to fit
the observations of the kinematics of the absorbers, the width 
$\sigma_i$ and the energy injection parameter $W_+$ ($f_{HI}$ and 
$r_g$ are fit only to the independent observational data of the
HI column density distribution).
Nevertheless, the fact that the two parameters
are enough to match the kinematic data and that both the velocity 
widths and the covering factor agree with observations for the same
value of the energy injection rate (as shown in Figs. \ref{covfac} 
and \ref{SCDMoverE}) is reassuring.

\subsection{Feasibility of Spherical Halo Models}

PW claimed to have shown that spherical halo models were ruled out
by the observed kinematics of the absorption components, specifically
by the fact that the components often show an asymmetric distribution
of velocity. Our analysis here will yield a different result.
We show that spherical halo models do not have serious difficulty
passing the shape tests of PW, as long as enough freedom is allowed
in the distribution of halo circular velocities and the number of
clouds at different radii. We need to emphasize here that we are
not proposing that the real absorption systems are part of
{\it exactly} spherical halos. A reasonable presumption is that
the gas in the damped absorbers is supported partly by rotation and
partly by random motions within a spheroidal halo, with rotation 
possibly
becoming more important in the center, where a disk may form.
But given the present observations, we shall show that the simple
spherical model is still consistent with them and is useful 
to capture the essential physical properties of the absorbers.
This suggests that it may be difficult to find an unambiguous 
signature of rotation using only absorption spectra in single lines
of sight.

  The statistical tests we use are fully described in PW, but we shall 
summarize them here. First, one defines the velocity width $\Delta v$ of
every absorber as that of the interval containing 90\% of the column
density of the absorber (leaving 5\% on each side; here column density
refers simply to the integrated optical depth of the low-ionization
absorption line being used). The Mean-Median test statistic is
\begin{equation}
f_{mm}=\frac{|v_{mean}-v_{med}|}{(\Delta v /2)} ~,
\end{equation}
where $v_{mean}$ is the midpoint of the velocity interval, and
$v_{med}$ is the median velocity (i.e., the point which has half the
column density on either side).
The Edge-Leading test statistic is
\begin{equation}
f_{edg}=\frac{|v_{pk}-v_{mean}|}{(\Delta v /2)} ~,
\label{dvw}
\end{equation}
where $v_{pk}$ is the velocity with the maximum optical depth (i.e.,
the first peak, as defined previously).
The Two-Peak test statistic is
\begin{equation}
f_{2pk}= \pm \frac{|v_{pk2}-v_{mean}|}{(\Delta v / 2 ) } ~,
\end{equation}
where $v_{pk2}$ is the velocity of the second peak, and the
negative sign is used if the second peak is on the opposite side of
the mean velocity from the first peak. 
The second peak is defined as previously. As we have seen, very often
there is no second peak, and in this case the location of the first
peak is used in its place. 

In Figure 5 of PW2, Kolmogorov-Smirnov (KS) test probabilities for
their different statistics and models are shown.  Their ``IH" 
(isothermal halo) model is ruled out by the velocity test and the
``two-peak" test.  However, the results of the same tests for our 
models, given in Table 2, indicate that
\begin{table}
\begin{tabular}{lcccccc}
\tableline\tableline
Model & $\sigma_8$ & $W_+/W_0$ & Velocity & Mean-Median & Edge-Leading & 
Two-Peak  
\\ \tableline
SCDM & 0.7 & 0.6 & 0.4\% & 24\% & 2\% & 81\%   \\ 
SCDM & 0.7 & 1.2 & 52\% & 34\% & 23\% & 29\%   \\ 
SCDM & 0.7 & 1.8 & 97\% & 29\% & 20\% & 19\%   \\ 
$\Lambda$CDM & 0.8 & 1.2 &43\% & 51\% & 33\% & 34\%   \\ \tableline
\end{tabular}
\label{KStab}
\caption{KS probabilities using the shape statistics
of Prochaska \& Wolfe.}
\end{table}   
the simulated lines are consistent with the observed ones based on all
the tests used by PW.
So why does the model of PW fail the test? The velocity test
obviously fails because they chose to give all of their halos an
internal velocity dispersion of 190 $\kms$, and then fixed the number of
components to ten. Thus, the velocity width was generally close to the
interval containing 90\% of the column density in a Gaussian profile,
$3.4 \times 190 \kms \sim 600 \kms$. 
The assumptions they made were too restrictive to reach any general
conclusion on the feasibility of a spherical model.

\begin{figure}
\centerline {
\epsfxsize=4.8truein
\epsfbox[70 32 545 740]{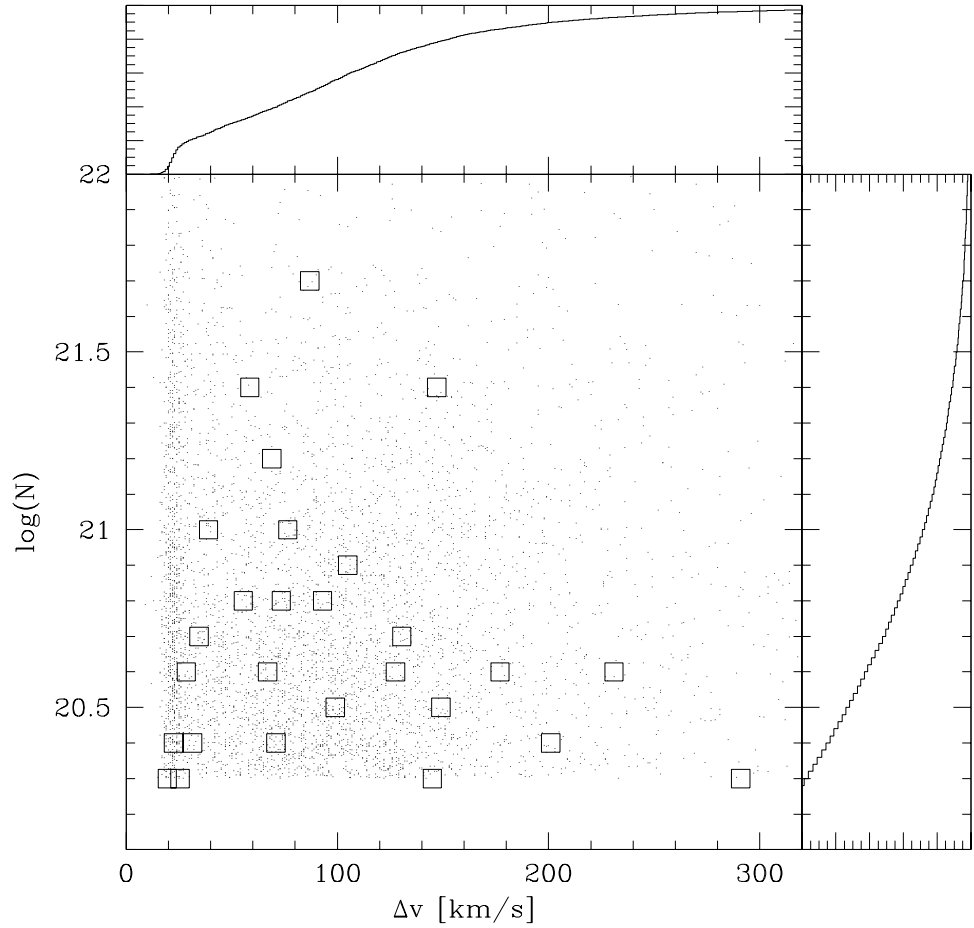}
}
\caption{Dots show column density vs. velocity width for a set of 
5400 simulated systems (SCDM, $\sigma_8=0.7$, $W_+=
1.8W_0$).
The squares show the 27 systems of sample C of PW2. 
The top and side panels show the cumulative distributions. }
\label{vvN}
\end{figure}

  Another comparison can be made between the joint distributions of 
column density and velocity in our model and in the observations.  
Figure \ref{vvN} shows our model predictions for the SCDM model along
with the observed systems of PW.
This figure can be compared directly to Figure 16 of PW2.  
The 2-dimensional KS test probability for this comparison
is 42\%. The distributions of column densities and velocity widths
are approximately independent in the model and the observations.
Attempts to model these systems in the future must keep this 
comparison in mind.  For example,  a model with a thin but very
dense disk surrounded by a halo of low column density clouds could
produce an anti-correlation between column density and velocity
width.  A line of sight that hit the disk would be dominated
by the one high column density component, while lines of sight which
only passed through the halo would have lower column density with
multiple components and a large $\Delta v$.

  Haehnelt, Steinmetz, \& Rauch (1998) found that absorption profiles
drawn from hydrodynamic simulations of galaxy formation also pass the
tests of PW (a preliminary comparison suggests that distinguishing
our model from their simulations using these shape tests would  
require many more observed systems).  
They found that their structure was produced by a
mixture of rotation, random motion, infall, and merging.
While the morphology of the multiple absorption components is of
interest, we agree with their conclusion that these tests of the
details of the velocity structure are not currently very constraining.
We also emphasize their point that KS probabilities must be used with
care. They rule out only the {\it specific} model they are applied
to, including any details of the model that are unrelated to the
main question of the presence or absence of rotation. 
The KS probabilities for tests of the kind
we have used here should be sensitive to changes we could make in the
details of the model such as the distribution of cloud column 
densities
at a given radius, a distribution of internal velocity dispersions,
spatial correlations of clouds, etc.

\subsection{ Absorption Velocity Widths: Constraints on the Power
Spectrum}

  Previously, lower limits on the amplitude of the power spectrum were
set from the total gas content of the damped \lya systems, using the
fact that the photoionization of the gas accreting on galaxies leads us
to expect that only halos above a certain mass can harbor damped \lya
systems (see Efstathiou 1992, Thoul \& Weinberg 1996, Navarro \&
Steinmetz 1997). Figure \ref{OHI} shows examples of the traditional
constraints (e.g., Mo \& Miralda-Escud\'e 1994) on the normalization of
the power spectrum for the models in Table 1 using the observed
$\Omega_{HI}$ at different redshifts. Now we can use the observations of
the velocity widths to set more stringent limits, since we have a better
knowledge of the internal velocities of the halos where the collapsed
gas is contained. Of course, the velocity widths depend on the internal
structure of the cloud model, but they are highly constrained by the
underlying distribution determined by the dark matter.    

  We found previously that without energy injection, no model can
reproduce the observed velocity widths. Figure \ref{SCDMoverE} shows the 
fraction of realizations of the 27 observed systems which display a
median velocity width in excess of the observed $77\kms$ for the 
SCDM model. Each of the
plotted points was computed with at least 200 realizations. We take the
energy injection rate at which the model produces the observed median
velocity width in less than 5\% of the realizations to be ``ruled out",
though of course it is only ruled out in the sense that we have defined
it.

\begin{figure}
\centerline {
\epsfxsize=4.8truein
\epsfbox[70 32 545 740]{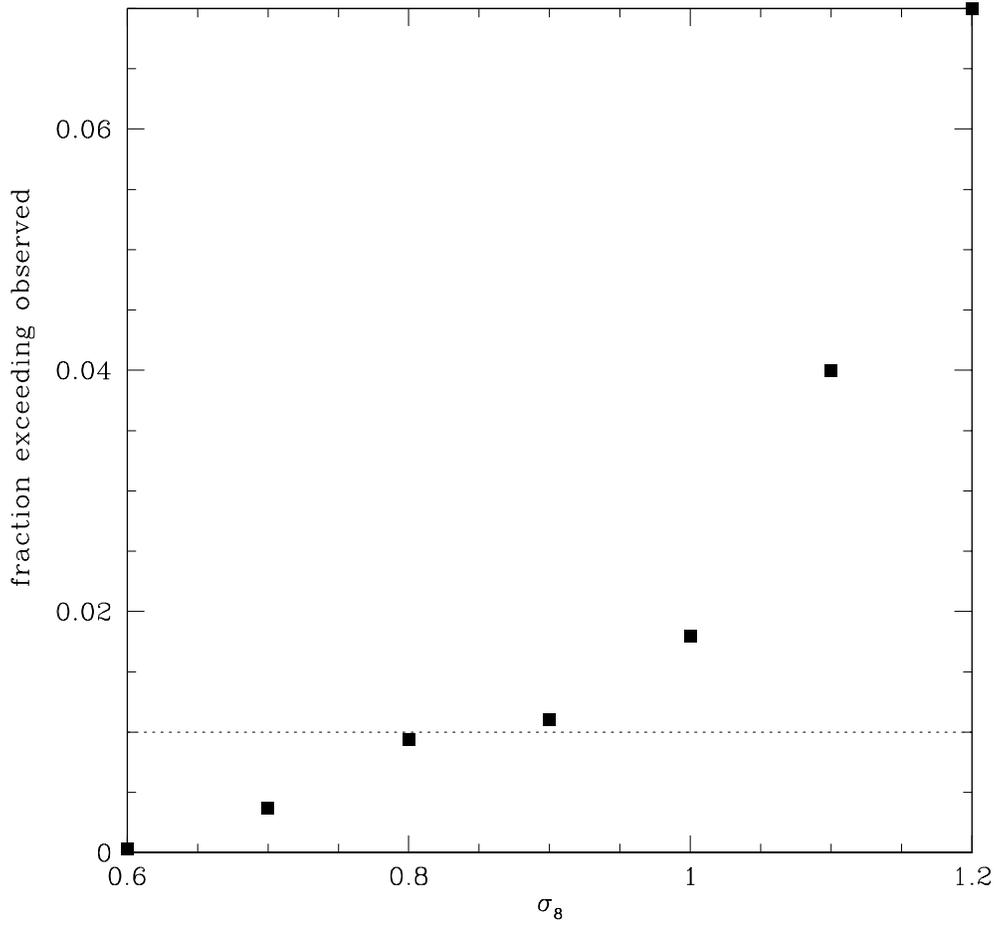}
}
\caption{Probabilities for finding median $\Delta v>148$ for the 
five highest redshift systems
(for the $\Lambda$CDM model with energy injection rate 
$1.8W_0$).}
\label{top5}
\end{figure}

  The ideal test to discriminate among cosmological models is not
necessarily to compare the distribution of velocity widths of all the
systems over a wide redshift range, because the models predict that the
halos with high velocity dispersion are particularly rare at high
redshift. Thus, a few observed systems with high velocity at the highest
available redshifts may provide a particularly stringent constraint. 
In Figure \ref{top5}, we plot the probability that the median velocity
width of the five highest redshift systems
($z=3.736$, 3.859, 4.08, 4.203, and 4.383) is greater than the observed
value, $148 \kms$, for the $\Lambda$CDM model as a function of its
normalization.
Due to the small number of systems, and the {\it a posteriori}
nature of the test, we say that the model is ``ruled out" only if
the probability for this test falls below $1\%$.
For this model we find that the lower limit to the power spectrum
normalization that satisfies this test is $\sigma_8 < 0.8$.
This is a strong test because the original $\Omega_g$ test is at the
highest possible redshifts ($z>4$). When a model is
close to failing the $\Omega_{g}$ test, the halo distribution is
dominated by the smallest halos, producing low velocity widths.

\begin{figure}
\centerline {
\epsfxsize=4.8truein
\epsfbox[70 32 545 740]{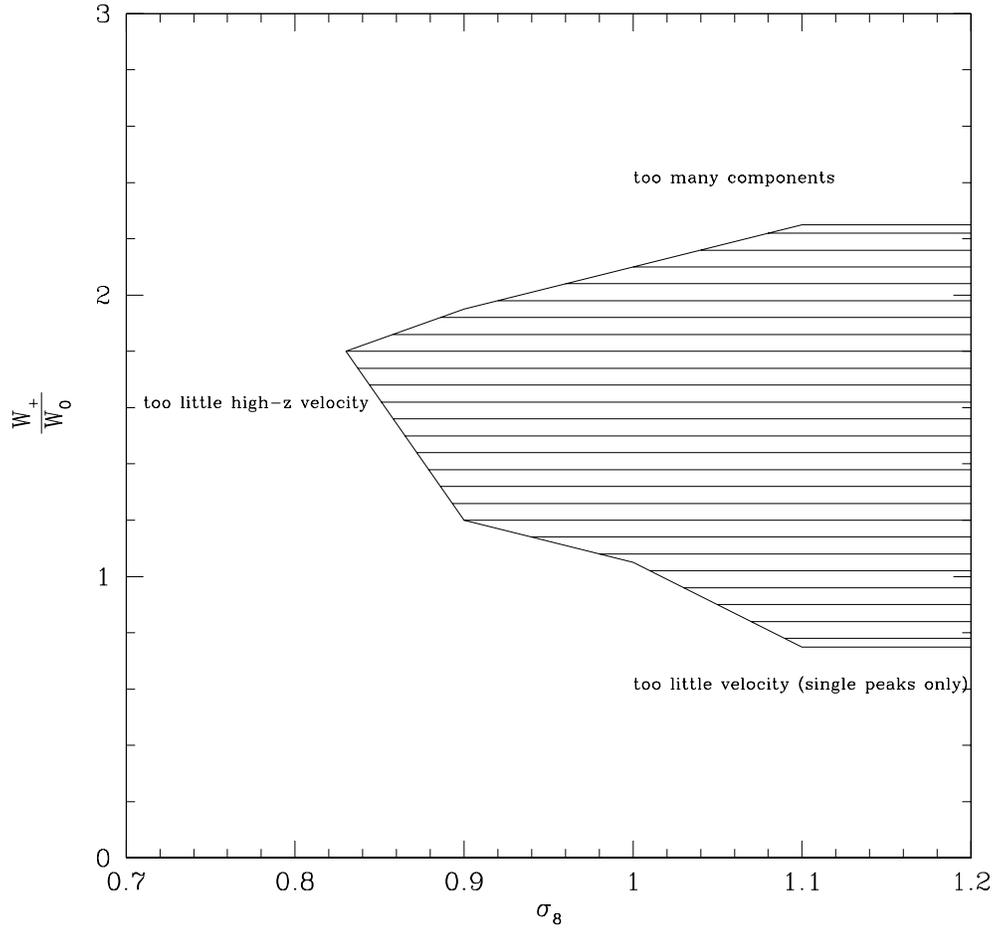}
}
\caption{For the $\Lambda$CDM model, the shaded region is allowed by 
the 
requirements that the model has the correct number of peaks, and
can produce the observed velocity widths.  The upper boundary comes
from overproduction of second peaks, the lower boundary is from
insufficient median velocity, and the left boundary is from insufficient
high redshift velocities.}
\label{fnewcon}
\end{figure}

  We have analyzed all the models in Table 1, generating 200
realizations of the 27 systems and computing the probability of
obtaining the median velocity width and the number of multicomponent
lines of the observations. For every model, the highest normalization
which is ``ruled out" is determined from one of three conditions: either
the model must yield a probability less than 0.05 to exceed the observed
median $\Delta v$ of all 27 systems, or less than 0.01 to 
exceed the median $\Delta v$ of the
five highest redshift systems, or less than 0.05 of
having {\it no more} than 10 of the 27 profiles with a second peak.  
Figure \ref{fnewcon} shows the resulting constraints for the 
$\Lambda$CDM model.  Constraints are obtained on both the energy 
injection rate
and the power spectrum normalization $\sigma_8$.  
The upper constraint on $W_+$ is derived from the requirement that the 
model can produce
enough profiles without a second peak (there are 17 of these in the
data), and the lower constraint originates from the requirement
that the model can produce the observed median velocity width.
The lower bound on $\sigma_8$ is in practice the result of the 
requirement that the high redshift velocities can be produced.
We find that the constraints on the energy injection rate do not 
depend strongly on the cosmological model.  

The constraints we find on $\sigma_8$ for each model are listed in
Table 3.
\begin{table}
\begin{tabular}{lcc}
\tableline\tableline
Model & $\sigma_8$ & $\sigma(HR=100\kms, z=4)$ 
\\ \tableline
SCDM & 0.5 & 0.78    \\ 
$\Lambda$CDM & 0.8 & 0.79   \\ 
OCDM & 0.6 & 0.79  \\ 
MDM & 0.95 & 0.80   \\ \tableline
\end{tabular}
\label{constrainttab}
\caption{Lower limit to $\sigma_8$ for each model in Table 1, and the 
corresponding value of the {\it rms} fluctuations on spheres of radii
$HR = 100 \kms$, at $z=4$,
the median redshift of the five highest redshift systems.}
\end{table}    
We have listed the values of $\sigma(HR=100\kms, z=4)$ to 
show that the difference
between models is accounted for by the difference in fluctuation
amplitudes on the typical scale of the halos which can produce
the high redshift velocities in each model. Because the lower 
bound on $\sigma_8$ comes from the high redshift test, we use 
roughly the median $\Delta v = 148\kms$ of the
the five highest redshift systems (median $z\simeq 4$), which 
corresponds to $HR=87\kms$.
This shows that {\it in the context of our absorber model},
$\sigma(HR= 100\kms, z=4)\geq 0.78$ is a lower bound on the 
power spectrum
normalization which is fairly independent of the cosmological model.

\subsection{Impact Parameters}

  Our simple model of spherical halos of clouds for the damped \lya
systems makes a clear prediction for the impact parameters of the
absorbers from the centers of their host halos, once the core radius
parameter $r_g$ is determined from the observed incidence rate and the
covering factor of the clouds is known to be in agreement with
observations.
Essentially, this prediction is a consequence of the abundance of halos
predicted by the Press-Schechter model, and our assumption that the
high column densities associated with the damped systems originate in
the centers of these halos.

\begin{figure}
\centerline {
\epsfxsize=4.8truein
\epsfbox[70 32 545 740]{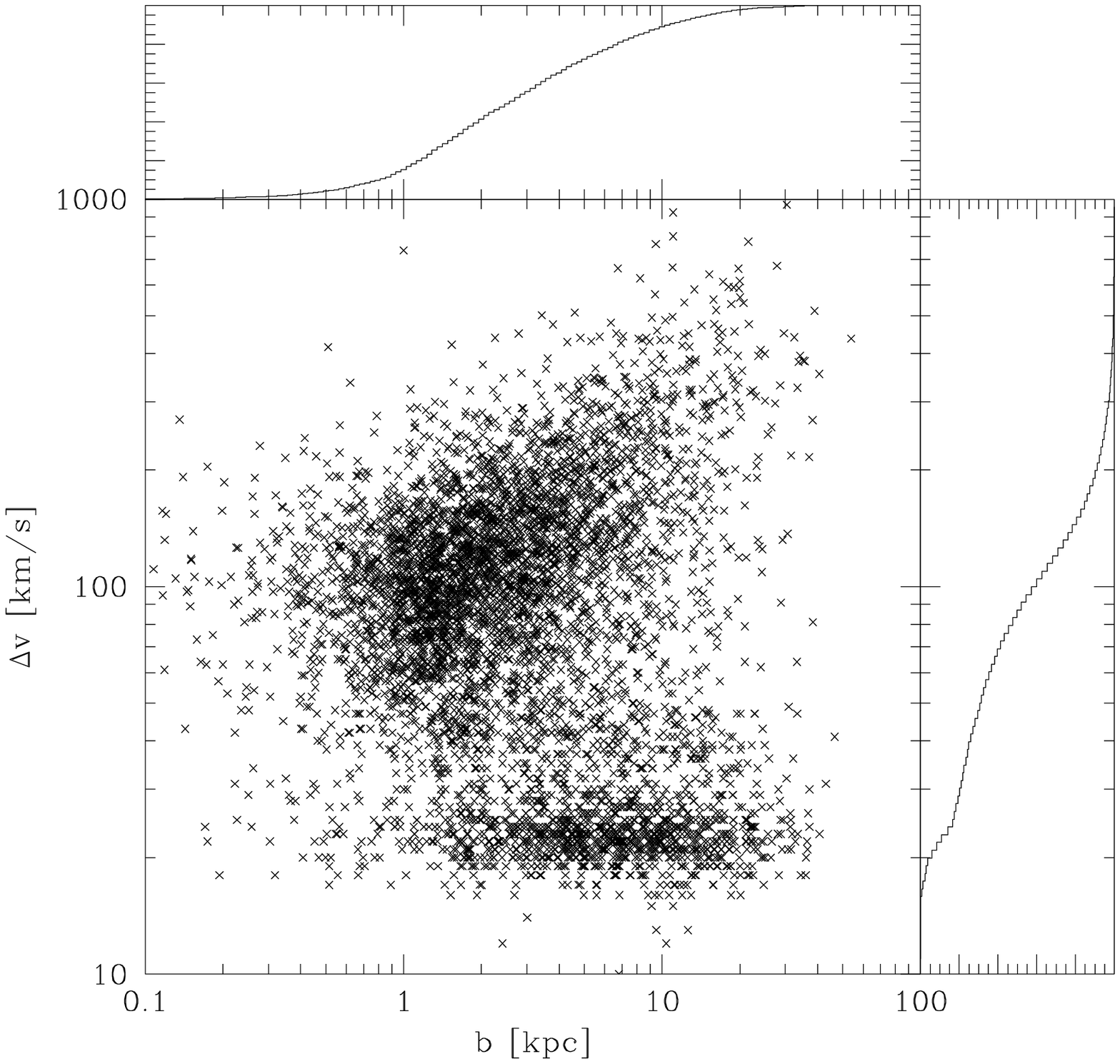}
}
\caption{Impact parameter vs. velocity width for a set of 5400
simulated systems (SCDM, $\sigma_8=0.7$, $W_+=1.8W_0$). 
The top and side panels show the cumulative distributions. }
\label{bvv}
\end{figure}

  We show in Figure \ref{bvv} a scatter plot of the velocity width and
the impact parameter for 5400 simulated absorption systems,
for the SCDM model with $\sigma_8=0.7$ and $W_+=1.8 W_0$.
The two adjacent panels on the right and top of the figure show the
cumulative distribution of the two variables separately.
We notice first of all that we predict that most of the systems have 
very small impact parameters, of order a few Kpc. This is in contrast 
to the well known result that very large sizes are required when damped 
systems are assumed to originate in spiral galaxies with the present 
abundance (e.g., Lanzetta \etal 1991). 
Small sizes are required in our model
because the majority of absorption systems arise in low-mass halos
with $V_c \lesssim 100 \kms$, and at high redshift these halos are
much more abundant than the present luminous galaxies. A halo with
$V_c=50 \kms$ at $z=3$ has a virial radius $r_{vir}\sim  10 \kpc$, and
as we discussed at the beginning of \S 2, the gas does not need to
collapse by a large factor in one of these halos after virialization
to reach the column densities of damped systems. Thus, the small impact
parameters are largely a consequence of the small sizes of the host
halos.

  The predicted parameters in the numerical simulation of Katz \etal
(1996) were slightly larger than ours. One of the reasons should be 
that halos with $V_c \lesssim 100 \kms$ were not resolved in this
numerical simulation.
Our predictions appear to be consistent with the observed impact 
parameters compiled by Moller \& Warren (1998).

  Figure \ref{bvv} also shows there is a long and substantial tail of
systems with large impact parameters. These tend to have small
velocity widths. In our model, the high impact parameter systems arise
from massive halos with large virial radii. However, at high impact
parameter the mean column density is low, implying that clouds reaching
the threshold column density of damped absorption have a very small
covering factor, and are therefore seen most often as individual
absorption components; hence their small velocity widths. It needs to
be pointed out here that the detailed distribution of impact parameters
is not to be considered a robust prediction, since it depends on our
choice for the mean gas profile in equation (\ref{denseprof}).

\begin{figure}
\centerline {
\epsfxsize=4.8truein
\epsfbox[70 32 545 740]{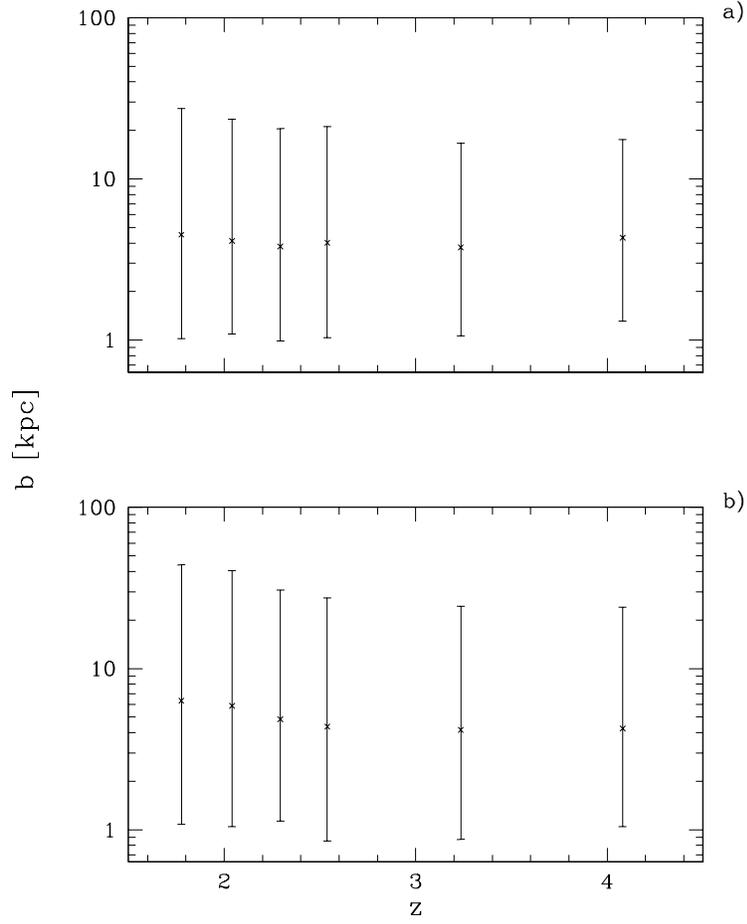}
}
\caption{Impact parameter vs. redshift for simulated systems.
The ``x" marks the median impact parameter, and the error bars
exclude 5\% of the generated systems at each extreme.  (a) and (b)
are $\Lambda$CDM with $\sigma_8=0.8$ and $\sigma_8=1.1$, respectively
(both have $W_+=1.8W_0$).}
\label{fsizes}
\end{figure}

  The distribution of impact parameters is expected to vary with
redshift. At lower redshift, halos are larger due to the older age
of the universe and to the increasing velocity dispersions.
Figure \ref{fsizes} gives the expected evolution of sizes for the 
$\Lambda$CDM model. 
The sizes are indeed predicted to increase somewhat
between $z=2.5$ and $z=1.5$, but the change is small from 
$z=4$ to $z=2.5$. The reason is that, according to the results of
Storrie-Lombardi \etal (1996), the median column density of the
absorbers decreases with redshift from $z=4$ to $z=2$, implying that
our parameter $r_g/r_{vir}$
needs to increase with redshift over the same range (see Fig. 3).
This seems consistent with the simple idea that, at high redshift, the
gas in virialized halos should have a more extended distribution within
the virial radius, because the process of dissipation had not yet
proceeded very far.

\section{Conclusions}

  A model of the multiple absorbing components in damped \lya systems
has been presented, based on randomly moving clouds in spherical halos.
We have found that the model can account for the observed kinematic
properties reported by PW. 
Whereas PW claimed that their observations
required the presence of a rotating system, we find that a spherically
symmetric system cannot be ruled out. The distribution of the multiple
components in the damped \lya systems seems to match that expected
for a random system of clouds, with the strongest absorbing component
being located at a random position within the absorbing interval.
It must be noted that even for a random spherical system, the strongest
absorbing component may often be at one edge of the absorbing interval
because most of the absorption occurs in a few components; for example,
if there are two main components, the strongest one is of course always
at the edge.

  The main new concept we have introduced in this paper is the rate
of energy dissipation in damped \lya systems. The high level of
turbulence evidenced by the observation of multiple components with
large velocity dispersions along random lines of sight shows that
shocks must occur often in the absorbing gas. When the size of the
absorbing systems is constrained by the requirement of matching the
observed gas content and incidence rate of the damped absorbers in
the context of the Press-Schechter model, we find that the rate at
which the kinetic energy of the random cloud motions needs to be
resupplied to balance the energy loss in cloud collisions is very
high, and cannot be accounted for from the gravitational energy of
merging halos.

  Haehnelt \etal (1998) analyzed model predictions for the kinematics
of the damped systems using numerical simulations of structure
formation. They also found that the kinematic properties of the
absorption systems reported by PW were in agreement with their
simulation, where the motions were generally not dominated by rotating
disks. Their simulation did not include any source of heating in
addition to the gravitational collapse of structure (such as energy
from supernovae explosions). Nevertheless, their simulated absorption
profiles have multiple components with characteristics similar to what
is observed. Thus, the question that arises in view of our results is
how the turbulent energy of the gas in the damped systems of their
simulation can be maintained without being dissipated at a rate
much faster than it can be replenished by gravitational mergers.

  We do not have a clear answer to this question at this point. A
reanalysis of the simulation would be desirable to check if the
energy dissipation rate is indeed as fast as predicted. If it is not,
this may indicate that the problem is related to the limited
resolution of the simulation; if it is, one should then find out why
gravitational energy is being provided in the simulation at a much
faster rate than indicated by our analytical calculation in this
paper. The impact parameters of the absorption systems predicted
by Haehnelt \etal (1998) are similar to those in our model (see their
Fig. 6 for examples of the size and morphology of their simulated
absorbers). Therefore, the rate at which the kinetic energy of the
gas clouds is lost in shocks should be approximately the same in their
simulation as in our model.

  What could the physical mechanism be for maintaining the gas motions
in damped \lya systems? If the major source of energy is indeed the
merging of halos due to gravitational collapse of structure, as we have
assumed in this paper, then our calculation of the rates of energy
dissipation or of the energy released in mergers must be wrong by a large
factor. The rate of energy dissipation could be much lower if the size
of the absorbing systems were much larger than in the Press-Schechter
model of halos. The extent of the absorbing systems could be very large
if damped \lya systems existed in clusters, and several of them were
intersected along a line of sight. However, if these clusters had
overdensities corresponding to virialized systems, then they would
already be included in the Press-Schechter model as massive halos
(and the absorbing systems within them should then correspond to our halo
clouds), whereas if their overdensity is low it is hard to see how the
covering factor of the damped systems within them could be large. 
But there could be other reasons why the extent of the absorbers is
larger than in our model. For example, it could be that only a small
fraction of collapsed halos contain extended atomic gas capable of
producing damped systems, with most of the halos having their gas
concentrated in the center at column densities $N_H > 10^{22}\cm^{-2}$,
possibly in molecular form. The observed rate of incidence of the damped
systems would then require larger cross sections for the halos that contain
extended gas, and therefore longer orbital times and lower dissipation
rates.

  At the same time, the rate of gravitational energy released could be
increased if the atomic gas that is observed in the damped systems is
a small fraction of the baryons contained in collapsed halos, but is
undergoing most of the energy dissipation required to form galaxies.
Thus, if damped systems are produced only by a small fraction of
collapsed halos, these halos could precisely be those undergoing a
merger. During a large merger, large amounts of gas should be added to
halos, which should dissipate energy at a rate much faster than average
and should give rise to damped absorbers with a large cross section.
Moreover, in this case the gravitational energy available
is not only that of the observed gas in damped systems, but also that
of the denser, possibly molecular gas that may be hidden from view
due to its very high column density (and consequently small cross
section). This possibility is related to the uncertainties mentioned
in \S 3 after equation (\ref{masscon}).

  The other possibility is that an additional source of energy maintains
the turbulent motions of the gas clouds in the damped absorbers. As
we have discussed in the text, energy from supernova explosions is a
plausible candidate. Determinations of the evolution of the metallicity
in damped \lya systems (e.g., Lu \etal 1996)
may help determine if the supernova energy
associated with the production of metals is sufficient to provide the
required energy. It is possible, however, that the heavy elements
produced are not expelled to the gas observed in damped absorbers, but
remain in very high column density systems with molecular gas. This
may be hard to reconcile with the low metallicities of many damped \lya
systems, especially at high redshift (Lu \etal 1996).

  A good understanding of the kinematics of the gas in damped \lya
systems can lead to useful constraints on the power spectrum of density
fluctuations, through the requirement that halos with the observed
velocity dispersion of the gas are sufficiently abundant. Using our
spherical halo model, we have obtained the condition that the
linearly extrapolated rms fluctuation on spheres of radius
$HR\simeq 100 \kms$ needs to be greater than $0.75$ at $z=4$.
We believe this lower limit should be robust,
despite having been derived in the context
of our model only. Our spherical halo model should probably yield the
maximum line of sight velocity dispersion of gas clouds that could be
compatible with a given model of the population of dark matter halos.

  Galaxy formation is intimately connected to the process of the
dissipation of the gravitational energy of the gas in halos. Therefore,
the resolution of the problem of the source of power for the random
motions of the absorbing components in damped \lya systems should yield
one of the clues we need to understand how galaxies form. The degree of
rotational support of systems at different column densities is of course
another important probe to the formation of disks. So far, it is clear
that the kinematics of the absorbers are not consistent with pure
rotation in a thin disk; but, given our results and those of Prochaska
\& Wolfe, it seems difficult to distinguish a thick disk from a purely
spherical halo on the basis of absorbing profiles along random lines
of sight only. More detailed modeling should be useful here. But new
types of observations, such as measuring the relationship of the
kinematics to the impact parameter and velocity of associated galaxies,
or the difference between absorption profiles of high and low ionization
systems, will probably be required to measure the degree of rotation.

\end{document}